\documentclass[pdflatex,sn-nature,iicol]{sn-jnl}

\usepackage{graphicx}%
\usepackage{multirow}%
\usepackage{amsmath,amssymb,amsfonts}%
\usepackage{amsthm}%
\usepackage{mathrsfs}%
\usepackage[title]{appendix}%
\usepackage{xcolor}%
\usepackage{textcomp}%
\usepackage{manyfoot}%
\usepackage{booktabs}%
\usepackage{siunitx}%

\makeatletter

\renewcommand\normalsize{%
  \@setfontsize\normalsize{8.5bp}{10.5bp}%
  \abovedisplayskip 8\p@ \@plus2\p@ \@minus1\p@
  \abovedisplayshortskip \z@ \@plus3\p@
  \belowdisplayshortskip 3\p@ \@plus3\p@ \@minus3\p@
  \belowdisplayskip \abovedisplayskip
  \let\@listi\@listI}%
\normalsize

\def\sectionfont{\reset@font\fontfamily{\rmdefault}\fontsize{9bp}{11bp}%
  \bfseries\selectfont\raggedright\boldmath\MakeUppercase}%
\def\subsectionfont{\reset@font\fontfamily{\rmdefault}\fontsize{8.5bp}{10.5bp}%
  \bfseries\selectfont\raggedright\boldmath}%

\def\figurecaptionfont{\reset@font\fontfamily{\rmdefault}\fontsize{7.5bp}{9bp}\selectfont}%
\def\tablecaptionfont{\reset@font\fontfamily{\rmdefault}\fontsize{7.5bp}{9bp}\selectfont}%

\makeatother

\unnumbered

\begin{document}


\title[Filler restructuring in MAEs]{Filler restructuring and surface roughness in magneto-active elastomers with magnetically hard versus soft particles}

\author*[1]{\fnm{J\'ulio P. A.} \sur{Santos}}\email{julio.palma.de.assuncao.santos@univie.ac.at}

\author[2]{\fnm{Mehdi} \sur{Hasanzade}}

\author[3]{\fnm{Chaitanya} \sur{Doifode}}

\author[3]{\fnm{Raphael} \sur{Kriegl}}

\author[2]{\fnm{Alexander} \sur{Kovalev}}

\author[3]{\fnm{Mikhail} \sur{Shamonin}}

\author[2]{\fnm{Stanislav N.} \sur{Gorb}}

\author[1]{\fnm{Sofia} \sur{Kantorovich}}

\affil*[1]{\orgdiv{Faculty of Physics}, \orgname{University of Vienna}, \orgaddress{\street{Kolingasse~14-16}, \postcode{1090}, \city{Vienna}, \country{Austria}}}

\affil[2]{\orgdiv{Functional Morphology and Biomechanics, Zoological Institute}, \orgname{University of Kiel}, \orgaddress{\street{Am Botanischen Garten~1-9}, \city{Kiel}, \postcode{24118}, \country{Germany}}}

\affil[3]{\orgdiv{East Bavarian Center for Intelligent Materials (EBACIM)}, \orgname{Ostbayerische Technische Hochschule (OTH) Regensburg}, \orgaddress{\street{Seybothstr.~2}, \city{Regensburg}, \postcode{93053}, \country{Germany}}}


\abstract{
In magneto-active elastomers (MAEs), a magnetic field reshapes the surface and with it adhesion, friction and wetting.
Most use magnetically soft carbonyl iron, whose restructuring has lacked a particle-resolved description.
We put forward a molecular dynamics model with magnetizable particles and compare thin layers of hard and soft fillers.
Both converge to the same high-field state: a labyrinth of bundled field-aligned chains roughening the surface by about one particle diameter.
Hard fillers break and rotate in-plane chains above a threshold field, roughening the surface abruptly.
Soft fillers build chains particle by particle, roughening it continuously.
The model reproduces the magnetization of carbonyl iron MAEs above about \qty{250}{\milli\tesla} and the magnitude of their roughness.
Measured roughness varies linearly with inverse zero-field Young's modulus: matrix stiffness, not particle number, controls it.
At equal saturation magnetization, filler type sets not how rough the surface becomes, but whether it roughens abruptly or continuously.
}


\maketitle


\section{Introduction}\label{sec:intro}

Magnetically controllable soft materials are becoming increasingly desirable for modern applications in soft robotics and actuation~\cite{review-magnetic-soft-matter-and-robots, soft-robotic-review-2}, where an external magnetic field can change the shape of a functional element and the mechanical properties of the constitutive material. Magnetic soft materials often show rich and complex behavior due to the coupling between magnetic and elastic degrees of freedom, observed in magnetic brushes, gels, colloidal systems, and elastomers~\cite{magnetic-soft-polymer-composites,1995-shiga,2000-zrinyi,Stepanov2008,Wereley2014,Morillas2020,1996-jolly,1999-ginder,2000-carlson,2006-varga,2012-boczkowska,2016-odenbach}. Materials exhibiting such characteristics can be employed not only in the fabrication of adaptive damping systems, vibration absorbers, and compact manipulators and sensors, but also in the development of artificial muscles and scaffolds for cultivating biological tissues required in regenerative medicine and transplantology~\cite{2000-carlson,2006-deng,2013-li,2014-li}.

Magneto-active elastomers (MAEs) --- composite materials combining a rubber-like polymer network with embedded magnetic nano- or microparticles --- are particularly attractive in this context, as magnetic fields change not only their shape and bulk mechanical properties~\cite{shamonin-stiffness} (elastic moduli, plasticity) but also their surface properties~\cite{roughness-gasper,surface-roughness-kramarenko,li-roughness-increase-and-decrease,drotlefMagneticallyActuatedPatterns2014,kovalevMagneticallySwitchableAdhesion2022,krieglTunableContactAngle2023} (roughness, wetting, adhesion, and friction).
In most devices the surface governs performance, since adhesion, friction and wetting are all set there. What matters for design is therefore not only how rough a layer can become, but how that roughness is reached as the field is ramped --- abruptly at a threshold, or continuously.

That response is set by a materials choice made at the formulation stage: whether the filler is magnetically hard (MH), retaining a permanent moment, or magnetically soft (MS), acquiring a moment induced by the local field.
Almost all MAEs fabricated for applications use MS fillers, overwhelmingly carbonyl iron powder, because they are inexpensive, have a high saturation magnetization, and leave the composite non-magnetic at zero field.
Hybrid MAEs that combine both fillers can outperform either filler alone~\cite{moreno-hybrid-MAE}; a particle-scale description of the two pure limits is therefore a prerequisite for designing them.

How each filler restructures the MAE, however, has proved difficult to observe.
MAEs have largely been understood through experiments and empirical relations, with a focus on magnetostriction~\cite{2008-guang,Diguet2010,silva2022giant,Glavan2024,romeis-model-susceptibility-130}, changes in elastic moduli~\cite{2014-pessot,2016-pessot, Snarskii2021}, wetting~\cite{sorokin2018controllable,krieglTunableContactAngle2023,chen2021magnetic}, and surface roughness~\cite{roughness-gasper, zhangControllableMagneticRoughness2020, kriegl2022microstructured}. Computer tomography can generate static images of particles embedded in an elastomer, but is currently unsuitable for following the dynamics of structural transformations and is constrained by the particles' size and shape~\cite{Borbath2012, Schumann201788, Flakes_Malte_2017}.
Theoretically, MAEs are often described by macroscopic material models coupled to a continuous representation of the magnetic interactions and solved with finite element methods (FEM)~\cite{raikher2008shape,ivaneyko2012effects,brand2014macroscopic,Nadzharyan2018316,stolbov2011modelling}; these lack the spatial resolution needed to capture the single-particle-scale magneto-elastic coupling that drives the macroscopic changes, and the spatial ordering of the particles.

FEM studies that resolve individual particles do exist~\cite{metsch2016numerical, guan-fem-vs-ezp-sufrace-roughness}, including for magnetizable fillers, but the computational cost is high and the element distortion that accompanies chain formation limits the deformations they can reach: the chains and closely packed arrangements that are present in MAEs are beyond what the method can follow.
Particle-resolved modeling has otherwise concentrated on MH fillers, because the interactions between MS particles are considerably more complex, and the few works that extend beyond the dipolar approximation~\cite{biller2019mesomechanical,romeis2026beyond} address the magnetization itself rather than the particle arrangement it produces.
To reach particle-scale resolution while still representing a thin slab, S\'anchez \textit{et al.}~\cite{pedro-elastomers-surface,pedro-alla-inelastic-deformations,alla-flake-magneto-elastic-coupling,FORC_Dobroserdova_2020,Dobroserdova_2023_PRE_FORC_SFD,dobroserdova2026influence} introduced two molecular dynamics models in which the particles are explicit point dipoles and the elastomer is coarse-grained either as springs connecting each particle to a fixed point in space~\cite{pedro-alla-inelastic-deformations}, or as a non-linear network of springs connecting the particles to one another~\cite{pedro-elastomers-surface}. The first is intended for low particle concentrations ($\lessapprox 0.1$)~\cite{pedro-alla-inelastic-deformations,dobroserdova2026influence} and the second for higher concentrations ($\gtrapprox 0.2$)~\cite{pedro-elastomers-surface} used in devices.
For the filler actually used in applications, the field-driven restructuring of a dense layer and the surface relief it produces still lack a particle-resolved description, leaving the design question --- which filler gives which actuation profile --- without a computational answer.

Here we build on the model of S\'anchez \textit{et al.}~\cite{pedro-elastomers-surface} to include magnetically soft particles, and compare thin MAE layers built from magnetically hard particles (MHP) or magnetically soft particles (MSP) across particle volume fractions and matrix stiffnesses.
We find that both fillers converge to the same high-field microstructure --- a labyrinth of bundled, field-aligned chains --- but reach it along qualitatively different paths.
The model reproduces the measured magnetization of carbonyl iron MAEs close to saturation, captures the magnitude of their surface roughness, and preserves the observed trends at lower fields.
The filler therefore does not set how much the surface roughens, but the mechanism behind the roughening: MH layers roughen abruptly above a threshold field, MS layers continuously.


\section{Results}\label{sec:results}

\begin{figure}[ht]
    \centering
    \includegraphics[width=0.8\columnwidth]{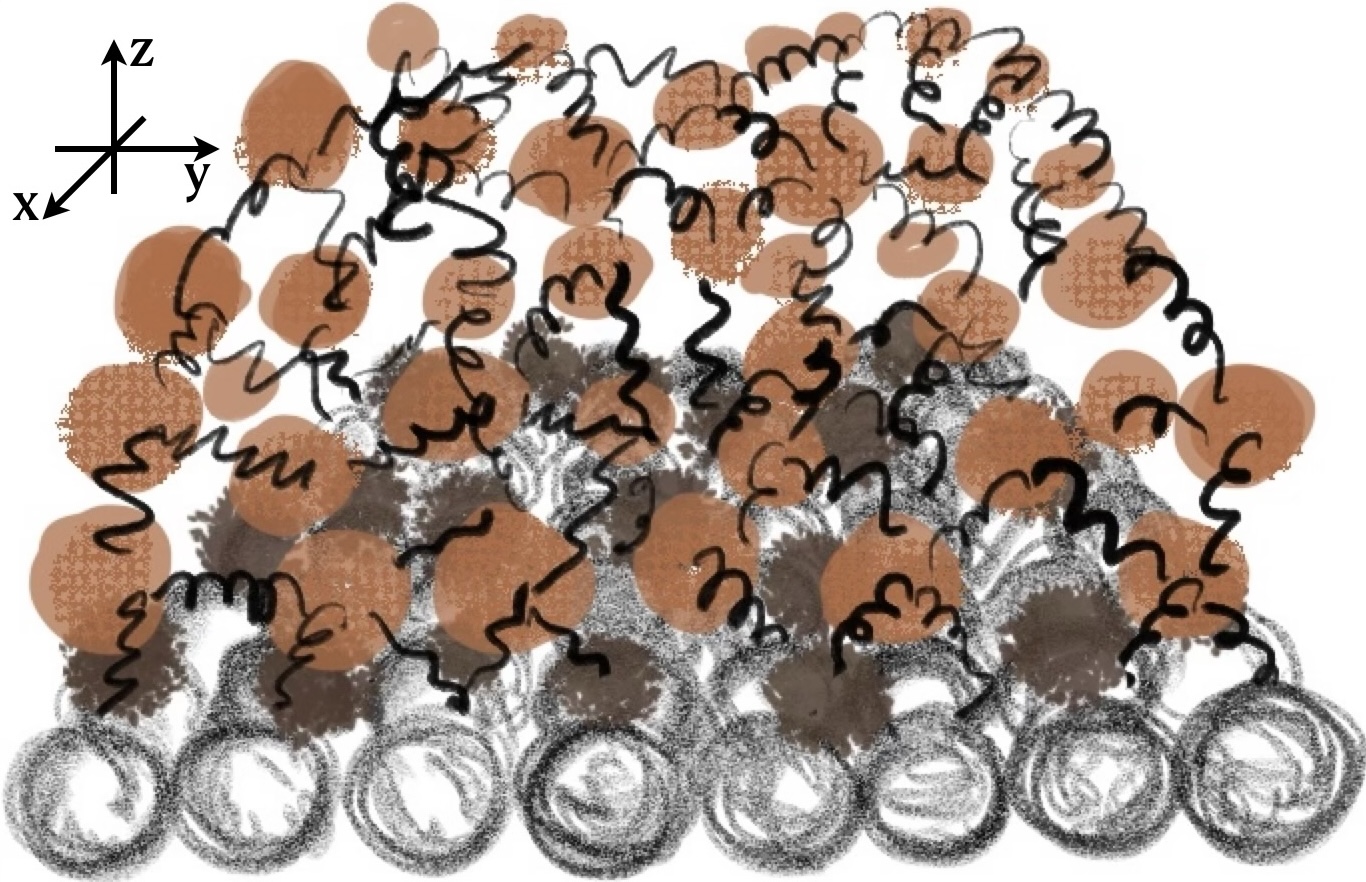}
    \caption{\textbf{Coarse-grained model of a magneto-active elastomer layer.}
    Magnetic particles (light and dark brown) are embedded in an elastic matrix represented by harmonic springs connecting the particles.
    The lowest layer of grey spheres is the substrate.
    All magnetic particles except those adsorbed on the substrate (dark brown) move in all three directions; the adsorbed particles move only within the substrate plane, with their $z$ coordinate fixed.
    The external magnetic flux density $\mathbf{B}$ is applied along $z$, perpendicular to the layer.
    Model and parameters are defined in the \textit{Methods}.
    }
    \label{fig: model-sketch}
\end{figure}

We model the MAE as a dense distribution of magnetic particles interconnected by springs~\cite{pedro-elastomers-surface} and adhered to a substrate (Fig.~\ref{fig: model-sketch}).
This lets us resolve the magnetic interactions at the particle scale while coarse-graining the elastic matrix into a network of springs --- orders of magnitude cheaper than a continuum description.
The model, the magnetization schemes for MH and MS particles, and the simulation protocol are given in the \textit{Methods}.
We simulated elastomer layers of thickness $10\sigma$ with magnetic particle volume fractions $\varphi\in\{0.20,0.25,0.30\}$ and elastic spring constant distributions $K_{*}\in\{K_{\text{low}}, K_{\text{high}}\}$, using four independent initial configurations for each case. Each configuration was run twice, once with magnetically hard particles (MHP) and once with magnetically soft particles (MSP).

Results are presented versus the non-dimensional field $\xi=\mu_{\infty}B/k_{\text{B}}\text{T}$ (Langevin parameter), calculated in simulation units with $k_{\text{B}}\text{T} = 1.61\times10^{-8}$ (see the \textit{Methods}).
A conversion table between the non-dimensional Langevin parameter and the magnetic flux density $B$ in SI and simulation units is given in Table~\ref{tab: xi-to-B}.

All simulations start from the configuration at $B = 0$ and are equilibrated directly at the target $B$ value; hysteresis effects are therefore not considered here.

Throughout, solid markers denote magnetically hard and open markers magnetically soft fillers; warm colors denote the stiff matrix and cool colors the soft one, while marker shape and color shade distinguish the volume fractions. For clarity the legend is given once per figure, in its final panel.

\subsection{Two distinct restructuring pathways}\label{sec:results-energies}

Although MAEs containing MHP and MSP reach similar high-field structures, they do so through fundamentally different particle restructuring mechanisms that originate from very different zero-field configurations.

In elastomers containing MHP, the particles initially form long stable chains lying horizontally within the layer (in-plane). As the applied field exceeds a threshold value, these chains progressively break apart and are reoriented toward the field direction (out-of-plane), giving rise to long chains spanning the layer thickness.

In contrast, elastomers containing MSP begin from a random distribution of non-magnetic particles at zero field. As the external field is applied, the particles become magnetized and gradually assemble into out-of-plane chains by attracting and stacking neighboring particles along the field direction.

\begin{figure*}[t]
    \centering
    \includegraphics[width=\textwidth]{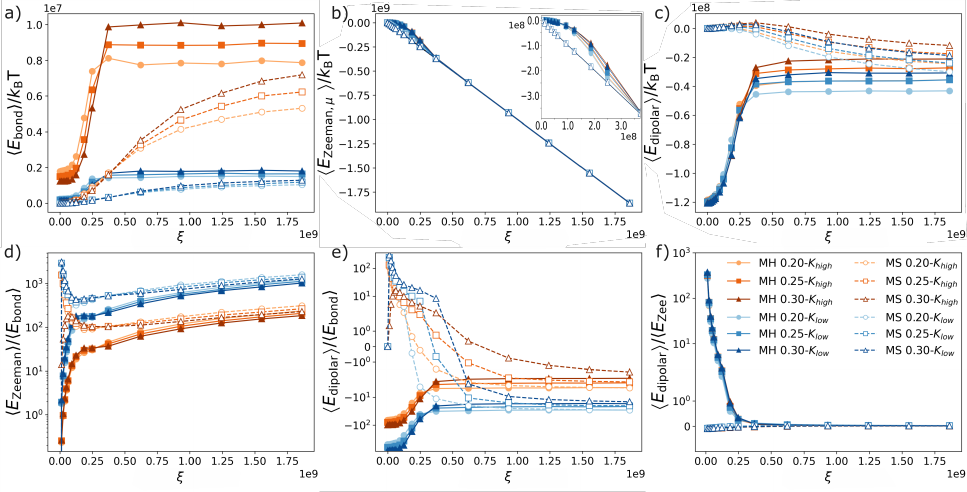}
    \caption{\textbf{Characteristic energies reveal two different paths to the high-field state.}
    Per-particle averages of \textbf{a} the elastic energy, \textbf{b} the $\mu$-normalized Zeeman energy, $\left<E_{\text{Zeeman}, \mu}\right> = -\tfrac{1}{N}\sum_{i} \hat{\boldsymbol{\mu}}_{i}\cdot\mathbf{B}$, and \textbf{c} the dipolar energy, as a function of the non-dimensional field $\xi=\mu_{\infty}B/k_{\text{B}}\text{T}$, for MAE layers with different volume fractions of magnetically hard (MH) or magnetically soft (MS) particles and different matrix stiffnesses.
    Ratios of \textbf{d} Zeeman to elastic energy, \textbf{e} dipolar to elastic energy, and \textbf{f} dipolar to Zeeman energy as a function of $\xi$; these ratios represent the relative contribution of the associated interactions to the equilibrium state.
    The inset in \textbf{b} is a zoom of the normalized Zeeman energy at low field.
    Solid markers denote MH and open markers MS fillers; warm colors denote the stiff matrix $K_{\text{high}}$ and cool colors the soft matrix $K_{\text{low}}$; marker shape and color shade denote the volume fraction.
    The legend for all panels is given in \textbf{f}.
    All energies are normalized by the thermal energy $k_{\text{B}}\text{T}$.
    Each point is an average over four independent initial configurations.
    }
    \label{fig: energies}
\end{figure*}

To understand the origin of the two mechanisms, we analyze three per-particle average energies (defined in the \textit{Methods}).
The elastic energy (Eq.~\eqref{eq: E_bond}) measures the elastic cost of the accumulated matrix deformation.
The $\mu$-normalized Zeeman energy (Eq.~\eqref{eq: E_Zee_mu}) measures the average alignment of the dipole moments with the applied field, with perfect alignment corresponding to $\langle E_{\text{Zeeman},\mu}\rangle = -B$.
The unnormalized Zeeman energy (Eq.~\eqref{eq: E_Zee}), used when calculating the ratios, measures how strongly the dipoles tend to align with the field.
Finally, the dipolar energy (Eq.~\eqref{eq: E_dip}) measures the dipole--dipole interactions between particles, with large negative values indicating head-to-tail configurations (most energetically favorable).
Figs.~\ref{fig: energies}~a--c show the evolution of each energy contribution with $\xi$.

The elastic energy (Fig.~\ref{fig: energies}~a) is lower for the less stiff MAEs (blue), reflecting weaker constraints on particle displacements. It is consistently higher for MHP than for MSP, indicating that MH elastomers deform more strongly than MS elastomers under the same applied field. Above a characteristic field strength, further particle displacement saturates, which is visible as a plateau for MHP, while MSP show a smoother continued trend toward saturation.

The $\mu$-normalized Zeeman energy (Fig.~\ref{fig: energies}~b) reflects the average alignment of magnetic moments with the applied field. MSP are, on average, aligned with the field for all field strengths, whereas MHP only exhibit significant alignment above a threshold field, as illustrated by the zoom inset.

The dipolar energy (Fig.~\ref{fig: energies}~c) highlights the most pronounced difference between MHP and MSP.
For MHP at $\xi=0$, it is close to $-1.2\times10^{8}$ and increases rapidly over a finite range above $\xi \approx 0.12\times10^{9}$, consistent with the rupture of initially favorable horizontal in-plane chains.
For reference, the non-dimensional per-particle energy of a perfect infinite chain is $\langle E_{\text{dipolar}}\rangle_{\text{chain}} / k_{\text{B}}\text{T} \approx -2.4 / k_{\text{B}}\text{T} \approx -1.5\times10^{8}$.
The zero-field configuration energy is therefore almost $80\%$ of that of an ideal infinite chain: the layer contains long, energetically favorable chains, which fall short of the ideal value because they are of finite length and not perfectly collinear.

For MSP, the dipolar energy is zero at $\xi=0$ (particles are not magnetized), becomes positive (net repulsive interactions) for $0<\xi \lessapprox 0.3\times10^{9}$ depending on particle concentration, and turns increasingly negative (net attractive interactions) for $\xi \gtrapprox 0.4\times10^{9}$, also depending on concentration.

This contrasting behavior highlights that, for MHP, the formation of out-of-plane (vertical) chains is initially unfavorable in terms of dipolar energy, whereas for MSP such vertical chain formation becomes progressively favorable beyond a threshold field, as magnetic moments are induced and begin to align. Despite these distinct pathways, both systems ultimately converge to the same dipolar energy at high fields, where MSP approach magnetic saturation and the two microstructures become effectively equivalent.

The ratios between the different energy contributions provide a clear picture of the field-induced competition between interactions that governs the system evolution under an applied field (Figs.~\ref{fig: energies}~d--f).

In Fig.~\ref{fig: energies}~d, we show the ratio of Zeeman to elastic energy.
The plot reveals that, for MH MAEs, this ratio exhibits only a very shallow maximum at $\xi \sim 0.2\times10^{9}$, which coincides with the onset of breaking of the horizontal chains (see Fig.~\ref{fig: energies}~f), where the ratio of dipolar to Zeeman energy drops sharply at nearly the same $\xi$. In contrast, this maximum becomes much more pronounced once the MS particles begin to acquire magnetization.

For MS elastomers in the absence of an external field, the particles are not magnetized and remain in their original equilibrium positions, resulting in $\langle E_{\text{bond}}\rangle = \langle E_{\text{Zeeman}}\rangle = \langle E_{\text{dipolar}}\rangle = 0$. As soon as a finite field is applied, Zeeman interactions become much larger than dipolar interactions, $|\langle E_{\text{Zeeman}}\rangle / \langle E_{\text{dipolar}}\rangle| \gg 1$, so particles primarily align with the field rather than forming chains. Chain formation thus sets in gradually: with increasing field strength, the particles acquire magnetization and the dipolar interactions intensify until they are sufficient to overcome the elastic constraints. This happens at $\xi \gtrapprox 0.6\times10^{9}$, with the precise value depending on particle concentration; see Fig.~\ref{fig: energies}~e, where the curves with open symbols cross $-1$.

\begin{figure*}[t]
    \centering
    \includegraphics[width=\textwidth]{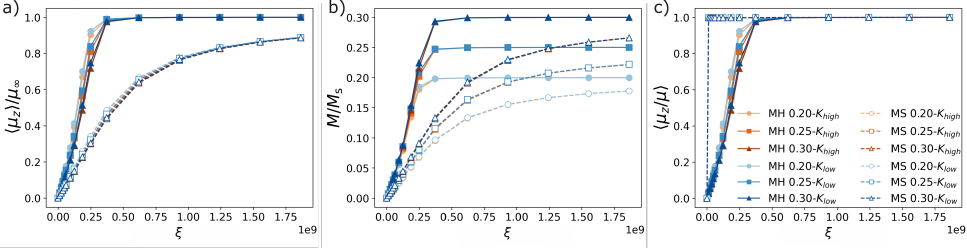}
    \caption{\textbf{Magnetization is rotation-limited for hard fillers and susceptibility-limited for soft fillers.}
    \textbf{a} Average $z$-component of the magnetic moment per particle normalized by the saturation moment, \textbf{b} magnetization normalized by the bulk saturation magnetization, and \textbf{c} average $\mu$-normalized $z$-component of the magnetic moment per particle, for increasing values of the non-dimensional field $\xi=\mu_{\infty}B/k_{\text{B}}\text{T}$.
    Solid markers denote magnetically hard (MH) and open markers magnetically soft (MS) fillers; warm colors denote the stiff matrix and cool colors the soft matrix; marker shape and color shade denote the volume fraction.
    The legend for all panels is given in \textbf{c}.
    Each point is an average over four independent initial configurations.
    }
    \label{fig: magnetization}
\end{figure*}

For MH elastomers subjected to weak fields, the particle arrangement is primarily governed by dipolar interactions, such that $|\langle E_{\text{dipolar}}\rangle / \langle E_{\text{bond}}\rangle| \gg 1$ and $|\langle E_{\text{dipolar}}\rangle / \langle E_{\text{Zeeman}}\rangle| \gg 1$ (Figs.~\ref{fig: energies}~e,~f). These interactions promote the formation of chains with maximal length. Since the MAE layer has a slab-like geometry, chains that lie horizontally within the layer are initially the most energetically favorable, as the lateral extent available for chain growth greatly exceeds the layer thickness.
As the field increases, Zeeman interactions become increasingly significant. For $\xi \gtrapprox 0.12\times10^{9}$, the in-plane chains are progressively broken and reoriented into field-aligned chains, while the elastic energy constrains their vertical extent. This results in an increasing number of out-of-plane chains with increasing field strength.

Overall, these ratios indicate that, as the applied field increases, the contributions from the Zeeman and elastic energies grow, while the influence of dipolar interactions diminishes.
Once the threshold field is surpassed, the behavior of the vertical chains is essentially governed by the competition between the elastic bonds, which tend to keep the MAE undeformed, and the applied magnetic field together with the dipolar forces, which promote deformation. This closely resembles the behavior of magnetic nanoparticles in ferrofluids and magnetic gels, which develop a Rosensweig instability when exposed to an external field oriented perpendicular to the fluid or gel surface~\cite{rosensweig85a,2007-bohlius}.

\subsection{Magnetic response}\label{sec:results-magnetization}

Consistent with the energy analysis, the average $z$-component of the magnetic moment per particle, normalized by the saturation moment, reaches the same saturation value for both MH and MS MAEs, but via distinct mechanisms.
In MH systems, the response is limited by the physical rotation of the particles and the competition between Zeeman and dipolar interactions, whereas in MS systems it is governed by the intrinsic susceptibility of the particles.

This difference is already visible in Fig.~\ref{fig: magnetization}~a, where MH elastomers reach saturation significantly faster than MS ones.
This is due to the relatively low initial susceptibility of the MSP ($\chi_{0} = 2.4$), which causes their magnetization to increase only gradually with field strength, while MH MAEs rapidly approach the maximum magnetization.

The saturation behavior is illustrated in more detail in Fig.~\ref{fig: magnetization}~b. The saturation magnetization, $M_\infty$, is determined solely by the particle volume fraction, the saturation magnetization of the material and the particle size, and does not depend on the elastic matrix or on whether the system is MH or MS. It can be written as
\begin{equation}
\frac{M_{\infty}}{M_{\text{s}}} = \varphi,
\end{equation}
where $M_{\text{s}} = \mu_{\infty}/V_{\sigma} \approx 1.91$ in simulation units and $\varphi$ is the particle volume fraction.

Finally, the difference in magnetization mechanisms is most clearly seen in Fig.~\ref{fig: magnetization}~c. The normalized $z$-component of the average magnetic moment of the MSP equals $1$ for all non-zero fields, indicating perfect alignment with the external field at all times.
For the MHP, in contrast, this quantity increases rapidly over a finite field range, reflecting the progressive reorientation of the particles.

The MHP are always magnetized to saturation but only fully align with the external field for $\xi \gtrapprox 0.35 \times 10^{9}$.
The MSP are always aligned with the field but become progressively magnetized as the field strength increases.

In the rotation-limited MHP, the elastic matrix affects how the system approaches saturation, because weaker mechanical constraints permit in-plane chains to reorient more readily. By contrast, the magnetization of MS elastomers depends only on the particle susceptibility and does not vary with the stiffness of the elastic matrix, which explains why the curves for both matrix stiffnesses (blue and red) coincide.

\subsection{Surface roughness}\label{sec:results-surface}

\begin{figure*}[t]
    \centering
    \includegraphics[width=\textwidth]{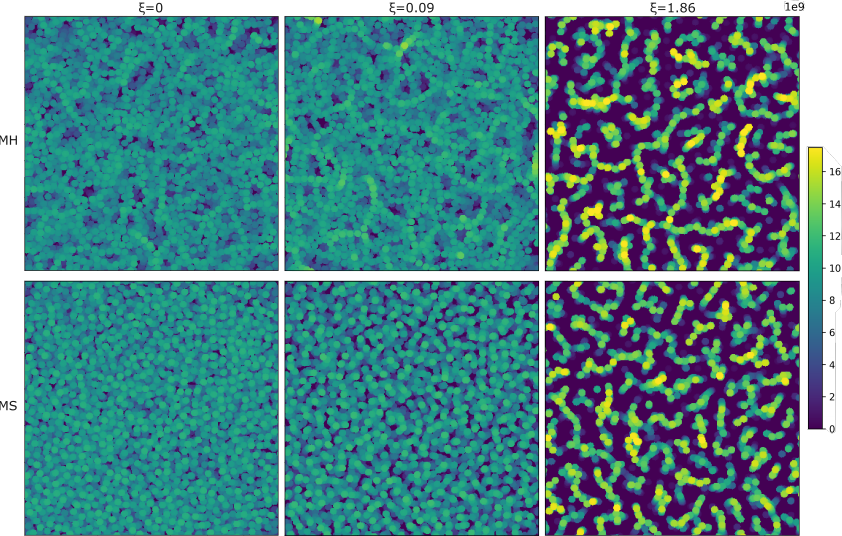}
    \caption{\textbf{Height maps of the free surface at three representative fields.}
    Height maps of the magnetic particles in magneto-active elastomer (MAE) layers adhered to a substrate, seen from above the free surface, for MAEs with magnetically hard (MH, top row) or magnetically soft (MS, bottom row) particles, with a layer thickness of $10\sigma$, volume fraction $0.20$ and the soft matrix $K_{\text{low}}$, at non-dimensional fields $\xi\in\{0, 0.09, 1.86\}\times10^{9}$, where $\xi=\mu_{\infty}B/k_{\text{B}}\text{T}$.
    Colors represent height according to the bar on the right, and the visible circles are individual magnetic particles.
    A height of $0$ is directly above the substrate.
    Length scales are in units of the particle size $\sigma$.
    }
    \label{fig: height-histograms}
\end{figure*}

\begin{figure*}[t]
    \centering
    \includegraphics[width=\textwidth]{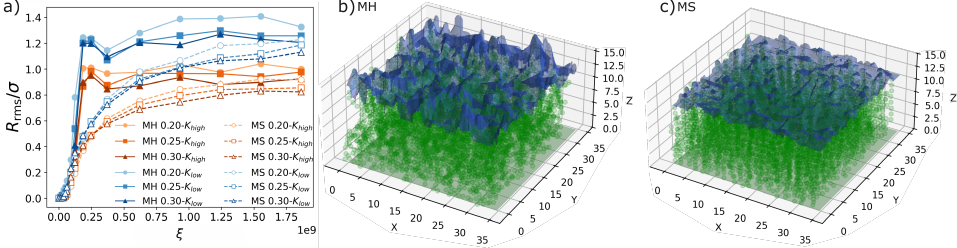}
    \caption{\textbf{Hard fillers roughen abruptly, soft fillers roughen gradually.}
    \textbf{a} Root-mean-square roughness parameter divided by the particle size, $R_{\text{rms}}/\sigma$, of an ITIM-estimated surface as a function of the non-dimensional field $\xi=\mu_{\infty}B/k_{\text{B}}\text{T}$, for magneto-active elastomers with magnetically hard (MH) or magnetically soft (MS) particles at different volume fractions and matrix stiffnesses.
    Solid markers denote magnetically hard (MH) and open markers magnetically soft (MS) fillers; warm colors denote the stiff matrix and cool colors the soft matrix; marker shape and color shade denote the volume fraction.
    Surface maps estimated using ITIM for \textbf{b} an MH elastomer and \textbf{c} an MS elastomer at $\xi=0.12\times10^{9}$.
    Length scales are in units of the particle size $\sigma$.
    }
    \label{fig: surface-roughness}
\end{figure*}

While the energy and magnetization analysis identifies which interactions dominate at various fields and which particle arrangements are energetically preferred, here we relate those findings to the actual particle configurations and surface topography, showing that MH and MS elastomers follow two distinct trajectories to reach similar high-field states.

Figure~\ref{fig: height-histograms} presents height maps of the MHP and MSP layers at three representative field strengths, illustrating the structural evolution discussed in the energy and magnetization analyses.

At $\xi=0$, the MHP assemble into extended in-plane chains, whereas the MSP remain randomly distributed, reflecting their initial field-free configuration.

At $\xi=0.09\times10^{9}$, some of the MHP chains are broken, with their free ends lifted by the magnetic field while portions of the original in-plane chains remain intact.
In contrast, the MSP aggregate with nearby particles to form numerous short out-of-plane chains distributed relatively uniformly throughout the sample.

At $\xi=1.86\times10^{9}$, both MHP and MSP consist exclusively of field-aligned out-of-plane chains. Rather than remaining isolated, these chains laterally aggregate into bundles that merge at junctions, producing an interconnected labyrinth-like network.
This corresponds to the physical picture proposed in Ref.~\cite{Snarskii2019}, in which an infinite cluster passing through the entire specimen is formed in a sufficiently high magnetic field.
The resulting structure resembles a maze of nearly parallel chain bundles linked by densely packed cluster regions where multiple bundles meet.

Figure~\ref{fig: surface-roughness}~a shows how the surface roughness (defined in the \textit{Methods}) varies with $\xi$.
The curves separate into three regimes, in agreement with the field structures in Fig.~\ref{fig: height-histograms}.
At $\xi=0$, the MAEs exhibit essentially no roughness.
Around $\xi\approx0.09\times10^{9}$, the roughness $R_{\text{rms}}$ of the MH elastomers increases sharply to large values, whereas the MS elastomers show a more gradual roughening.
For $\xi\gtrapprox0.18\times10^{9}$, the MH roughness reaches a plateau, and the MS roughness slowly approaches comparable values.
The abrupt increase in $R_{\text{rms}}$ for the MH elastomers is consistent with the energetic barrier associated with breaking the in-plane chains.
As expected and consistently across all data sets, a softer elastic matrix (blue) permits larger particle displacements and thus leads to rougher surfaces.

The volume fraction exerts a weak yet systematic influence: MAEs with lower particle loadings exhibit rougher surfaces, as also observed experimentally~\cite{roughness-gasper}.
The surface maps in Figs.~\ref{fig: surface-roughness}~b,~c, both at $\xi=0.12\times10^{9}$, illustrate the ITIM-based characterization of the surface topography.
For the MS elastomer, many short surface peaks are observed, whereas the MH elastomer displays only a small number of tall peaks, in agreement with the foregoing analysis and observations.

The sign of the field-induced roughness change is known to depend on the initial surface state: Li \textit{et al.}~\cite{li-roughness-increase-and-decrease} showed experimentally that initially smooth MAE surfaces roughen under field, whereas initially rough surfaces of the same material are smoothed, with both converging toward a similar intermediate roughness.
The field-induced smoothing recently reported by Guan \textit{et al.}~\cite{guan-fem-vs-ezp-sufrace-roughness} corresponds to the latter regime, as their samples are already rough at zero field.
Our layers start essentially flat and, thus, fall in the former regime: the field can only roughen the surface, by chain protrusion.

\subsection{Internal microstructure}\label{sec:results-microstructure}

\begin{figure*}[t]
    \centering
    \includegraphics[width=\textwidth]{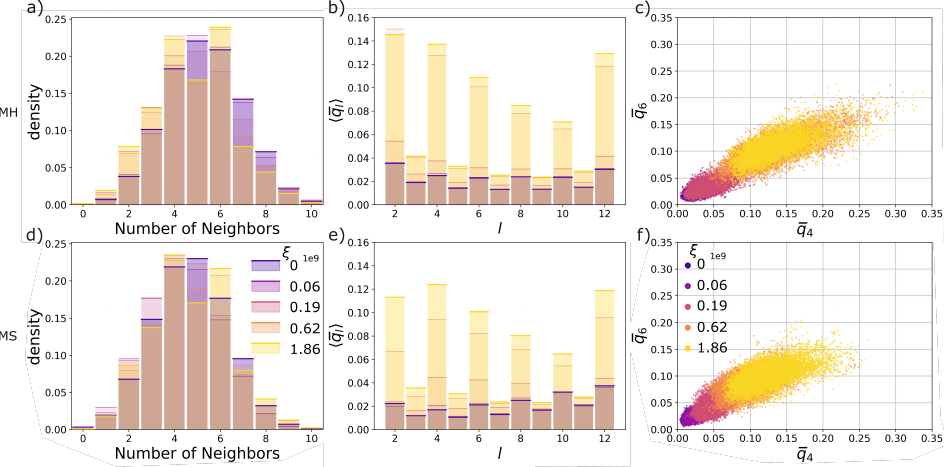}
    \caption{\textbf{Bulk restructuring resolved by coordination number and bond-order parameters.}
    Distribution of the number of neighbors per magnetic particle for \textbf{a} magnetically hard particles (MHP) and \textbf{d} magnetically soft particles (MSP) in magneto-active elastomers (MAEs).
    Average of the averaged bond-order parameters (BOPs) of the \textbf{b} MHP and \textbf{e} MSP.
    Joint distribution of the BOPs $\bar{q}_{4}$ and $\bar{q}_{6}$ for \textbf{c} MHP and \textbf{f} MSP.
    Colors correspond to the values of the non-dimensional field $\xi\in\{0, 0.06, 0.19, 0.62, 1.86\}\times10^{9}$, from dark (low field) to light (high field).
    Only particles in the bulk of the layer are included, identified from the plateau of the $z$-density profile (see the \textit{Methods}).
    Two particles are counted as neighbors when their center-to-center distance is smaller than $1.3\sigma$.
    }
    \label{fig: bulk-structure}
\end{figure*}

As discussed above, the MHP configuration exhibits a rapid change from initial in-plane chains to bundles of out-of-plane chains at high field, whereas the MSP configuration transforms gradually from a random arrangement to the same type of bundles. In both systems, the surface deformation is governed by a reorganization of the particle network in the bulk.

To characterize this reorganization, we examine the coordination-number distribution (Figs.~\ref{fig: bulk-structure}~a,~d), the average of the averaged bond-order parameters (BOPs, Eq.~\eqref{eq: BOP_avg}; Figs.~\ref{fig: bulk-structure}~b,~e), and the joint distribution of the averaged BOPs $\bar{q}_4$ and $\bar{q}_6$ (Eq.~\eqref{eq: BOP}; Figs.~\ref{fig: bulk-structure}~c,~f).
The coordination number describes the local connectivity between particles, while the BOPs measure the orientational symmetry of the local environment and thus differentiate disordered structures from locally ordered packings.

Consistent with the behavior observed for the energies and surface properties, at high fields (orange and yellow columns) the difference in the number of neighbors between MAEs with MSP and MHP is not significant.
In contrast, at zero and low fields, MHP exhibit substantially higher coordination numbers (violet and dark purple columns).
As the magnetic field increases, both systems evolve toward bundles composed of two or three nearly parallel chains, consistent with the labyrinth-like structures observed in Fig.~\ref{fig: height-histograms}. This structural motif explains the coordination-number distributions: a particle belonging to a single chain has two longitudinal neighbors, while adjacent parallel chains provide an additional two to four lateral neighbors, giving rise to the dominant populations around four and six neighbors in Figs.~\ref{fig: bulk-structure}~a,~d.
This observation by itself does not suffice to determine the internal structure of MAEs and must be supplemented by the BOPs.

At $\xi=0$, the two systems exhibit distinct particle organizations.
The MHP already form chains inherited from the field-free configuration, whereas the MSP remain nearly isotropic. This difference is reflected in the larger $\langle \bar{q}_{2}\rangle$ of the MHP, which captures the strong uniaxial character of the local environment.
The MSP display consistently smaller values of $\langle \bar{q}_{2}\rangle$, indicating the absence of a preferred local orientation (purple columns, Figs.~\ref{fig: bulk-structure}~b,~e).

The BOPs reveal that the high-field bundles possess a pronounced \emph{local hexagonal close-packed-like (hcp-like) orientational order}. This should not be interpreted as the formation of an hcp crystal; rather, the relative values of $\langle\bar{q}_{6}\rangle$, $\langle\bar{q}_{8}\rangle$, $\langle\bar{q}_{10}\rangle$ and $\langle\bar{q}_{12}\rangle$ indicate that neighboring chains adopt a locally hexagonal arrangement in the plane perpendicular to the chain direction. Such an arrangement is expected because magnetic interactions between neighboring chains favor staggered, closely packed configurations. As a consequence, the bundles develop a local hexagonal coordination in the transverse plane while remaining elongated along the field direction. The coexistence of large $\langle \bar{q}_2\rangle$ with hcp-like higher-order BOPs therefore reflects the anisotropic nature of the bundles: orientational order along the chain direction coexists with hexagonal packing of neighboring chains in the perpendicular plane.

The different transition pathways of MHP and MSP are most clearly visible in the $(\bar{q}_4,\bar{q}_6)$ distributions (Figs.~\ref{fig: bulk-structure}~c,~f). The MHP exhibit two well-defined populations corresponding to the initial in-plane chain network and the final bundled state, indicating that the restructuring proceeds through a relatively abrupt reorganization between two stable configurations. This bimodal distribution persists even when intermediate field values are included, suggesting that it is an intrinsic feature of the MHP transition rather than a consequence of the particular fields sampled. In contrast, the MSP display a continuous migration of probability between these two regions, consistent with the gradual field-induced assembly of chains into bundles. Correspondingly, the BOP spectra already develop weak hcp-like signatures at intermediate fields for the MSP, whereas the MHP retain the BOP characteristics of the initial chain network until the rapid transition occurs.

\subsection{Validation against measured magnetization curves}\label{sec:results-exp-mag}

\begin{figure}[t]
\centering
\includegraphics[width=0.75\columnwidth]{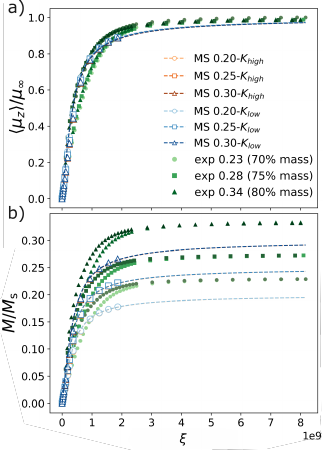}
\caption{\textbf{The magnetizable-particle model reproduces measured magnetization near saturation.}
\textbf{a} Average per-particle magnetic moment and \textbf{b} magneto-active elastomer (MAE) magnetization as a function of the non-dimensional field $\xi=\mu_{\infty}B /k_{\text{B}}\text{T}$, for experimental and simulation results on MAEs with different magnetic particle volume fractions.
Dashed lines are second-order modified mean-field (MMF2)~\cite{alexey-MMF2} fits to the simulation data, extending the curves into saturation.
A single horizontal rescaling, fixed by one anchor point, is applied to all volume fractions (see the \textit{Methods}).
Experimental data taken from Refs.~\cite{gasper-magnetisation,gasper-magnetisation-zenodo}.
}
\label{fig: mag_exp}
\end{figure}

To assess how physically meaningful the simulations are, we compare the computed quantities with experimental measurements.

We first validate the magnetizable particle model against the magnetization curves reported in Ref.~\cite{gasper-magnetisation}.
Subsequently, we contrast the simulated surface roughness with experimental measurements on elastomers of similar particle volume fractions subjected to comparable applied fields.

The elastomer is represented in a simplified form as a spring network, so matrix stiffness values that map directly onto experimental measurements cannot be extracted, and we do not convert the matrix stiffness into experimentally measurable units.

Figure~\ref{fig: mag_exp} shows that the magnetizable point dipole model reproduces the magnetic moment and magnetization curves of experimental MAEs~\cite{gasper-magnetisation}, even though these contain micron-sized multi-domain magnetic particles.
The average dipole moments are well represented for non-dimensional fields higher than $\xi \approx 1.5\times10^{9}$ --- $B\approx\qty{250}{\milli\tesla}$ --- with a single rescaling that is consistent across the different volume fractions.
Since this rescaling is a single horizontal shift fixed by one anchor point rather than a per-curve fit (see the \textit{Methods}), agreement at any other field is a genuine test of the model rather than a consequence of fitting.

For non-dimensional fields lower than $\xi \approx 1.5\times10^{9}$, we observe a systematic deviation: the experimental moments have a slower rise, as the multi-domain carbonyl iron particles have a low effective initial susceptibility due to domain walls and demagnetizing fields inside the particles.
These intra-particle effects are not included in the magnetizable dipole model, which leads to an over-magnetization of the simulated particles at lower fields.
Comparison between experiments and simulations at smaller fields should be done with this limitation in mind.

\subsection{Validation against measured surface roughness}\label{sec:results-exp-rms}

\begin{figure}[t]
\centering
\includegraphics[width=0.75\columnwidth]{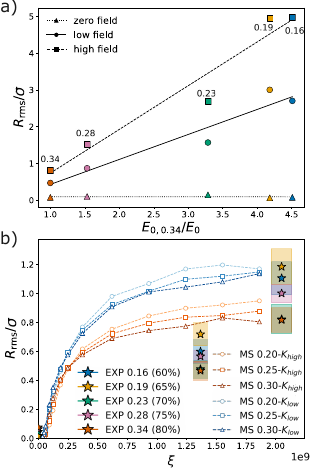}
\caption{\textbf{Zero-field stiffness, not particle number, controls the volume-fraction dependence of the roughness.}
\textbf{a} Linear relation between the root-mean-square roughness parameter divided by the average particle size, $R_{\text{rms}}/\sigma$, and the normalized inverse zero-field Young's modulus, $E_{0,0.34}/E_{0}$, shown at zero-, low- and high field, using the $\varphi=0.34$ sample as the reference modulus $E_{0,0.34}$.
Colors represent different volume fractions, labeled at the corresponding high-field point (squares).
\textbf{b} $R_{\text{rms}}/\sigma$ as a function of the non-dimensional field $\xi=\mu_{\infty}B/k_{\text{B}}\text{T}$ for simulations and experimental measurements on magneto-active elastomers (MAEs).
Orange and blue curves are simulations for MAEs with magnetically soft particles at different volume fractions and matrix stiffnesses.
Stars are experimental measurements for MAEs with different particle concentrations; boxes are the measurement error bars on both axes.
The experimental $R_{\text{rms}}/\sigma$ is rescaled to a common zero-field stiffness to allow direct comparison with the simulations, following the linear map in \textbf{a}.
}
\label{fig: exp-rms}
\end{figure}

Both in experiments and in simulations, the surface roughness increases with the applied field and reaches higher values for lower particle volume fractions (Fig.~\ref{fig: exp-rms}~b). This is especially pronounced for the experimental MAEs (Fig.~\ref{fig: exp-rms}~a).
In principle, the dependence on the volume fraction $\varphi$ can arise from two different effects.
On the one hand, $\varphi$ sets the number of particles available to chain or cluster: more particles can form out-of-plane chains protruding through the surface, but a denser packing also leaves the particles less free space, hindering the rearrangements required to form such chains.
On the other hand, $\varphi$ also sets the zero-field stiffness of the MAE, and at sufficiently high concentrations the rigid particles reinforce the elastic matrix~\cite{shamonin-stiffness}: a stiffer matrix further constrains the particle displacements needed to form out-of-plane chains, suppressing the surface roughening (as seen in the simulations, Fig.~\ref{fig: surface-roughness}~a).

For the experimental samples, the two effects can be weighed against each other by relating the roughness directly to the measured zero-field stiffness.
We find that $R_{\text{rms}}/\sigma$ follows an approximately linear relation with the inverse zero-field Young's modulus, $1/E_{0}$, normalized by a reference $E_{0,0.34}$ (Fig.~\ref{fig: exp-rms}~a).
We, therefore, conclude that the dominant driver of the roughness change between densities is not the number of particles available to form out-of-plane chains, but the effective change in the elastomer stiffness as the volume fraction varies.
The simulations reinforce this conclusion: increasing the particle concentration adds only a few extra harmonic springs, so the network stiffness at $\xi=0$ grows far more weakly than in experiments and, correspondingly, the roughness varies only weakly with the volume fraction.

We use this linear relation to rescale the $R_{\text{rms}}/\sigma$ curves onto a common zero-field stiffness, $R_{\text{rms}}/\sigma = (R_{\text{rms}}/\sigma)_{\text{source}} \times E_{0, \text{source}}/E_{0, \text{target}}$.
Since the rescaling is taken from a single linear map, the choice of reference is arbitrary: we take the experimental sample with $\varphi=0.34$ as the reference, as its roughness is closest to the simulation curves (Fig.~\ref{fig: exp-rms}~b).

After rescaling (Fig.~\ref{fig: exp-rms}~b), the $R_{\text{rms}}/\sigma$ curves display the same qualitative features for the simulated and experimental surfaces: the roughness is negligible at $\xi=0$, increases with the field, and is larger for MAEs with lower particle volume fractions. The order of magnitude is consistent across all results.
The clearest discrepancy is at intermediate fields, where the simulations have already reached their maximum roughness while the experimental elastomers have not.
We attribute this to the same limitation identified in the magnetization curves: the point-dipole magnetizable particles over-magnetize below $\xi\approx1.5\times10^{9}$ (Fig.~\ref{fig: mag_exp}~a), so the simulated particles restructure --- and roughen the surface --- at lower fields than the multi-domain carbonyl iron particles, whose lower effective low-field susceptibility delays this.


\section{Discussion}\label{sec:discussion}

We have implemented a particle-resolved molecular dynamics model of magneto-active elastomers (MAEs) with magnetizable fillers, and used it to compare thin MAE layers built from magnetically hard and magnetically soft particles across particle volume fractions and matrix stiffnesses.

Both MH and MS elastomers converge to the same high-field state --- a labyrinth of bundled, field-aligned chains --- but do so through two distinct paths: the MH in-plane chains, favored at zero field by the thin-slab geometry, are broken and rotated into alignment above a threshold field, whereas the MS chains are assembled one particle at a time from an initially random, unmagnetized arrangement, with Zeeman interactions dominating the dipolar ones at every field.
These two mechanisms manifest themselves in every observable we measured: the characteristic energies and their ratios, the magnetization curves, the surface roughness, the particle height maps and the local bond-order parameters.

The magnetization of MH layers is governed by the physical rotation of the particles, initially limited by the strong dipolar interactions, while that of MS layers is limited by the particle susceptibility.
The saturation magnetization itself does not depend on the elastic matrix or the magnetic nature of the filler; it is fixed by the particle volume fraction, the bulk saturation magnetization of the material and the particle size.
At high fields, both fillers show four to six neighbors per particle, consistent with bundles of two to three chains, together with an hcp-like local ordering of the chains.

Against experimental data, the magnetizable model reproduces the magnetization curves of carbonyl iron MAEs for fields close to saturation, $B\gtrapprox\qty{250}{\milli\tesla}$, captures the magnitude of their surface roughness, and preserves the same qualitative trends down to lower fields.
It also clarifies why the roughness decreases as the particle volume fraction increases.
Raising $\varphi$ acts in two distinct ways: it supplies more particles able to form out-of-plane chains, but it also stiffens the matrix, since the rigid inclusions reinforce the elastomer and restrict the displacements that chain formation requires.
The measurements separate the two effects, as $R_{\text{rms}}/\sigma$ varies linearly with the inverse zero-field Young's modulus at both low and high field, identifying the stiffening rather than the particle count as the dominant one.
The simulations provide the complementary control: there, added particles contribute only a few extra springs, the network stiffness at $\xi=0$ barely changes, and the changes in roughness are correspondingly much smaller.

These results are directly relevant to MAE-based soft-robotic and surface-engineering applications, in which the surface structure controls adhesion, friction and wetting, as the two restructuring mechanisms imply qualitatively different control characteristics.
In MH layers, the surface roughens only once the threshold field for breaking the in-plane chains is exceeded, and, because the broken chain segments rotate as objects of significant size, the intermediate-field surface is dominated by a few tall, sparsely distributed peaks; this is a switch-like response well matched to grip-and-release operations.
In MS layers, the roughness instead grows gradually and monotonically with the field, through many short and more uniformly distributed protrusions, offering continuous tunability of the surface texture.
The choice of filler particle is, therefore, not only a question of magnetic strength, but of the desired actuation profile: a threshold-like response with strong localized surface features near the threshold field, or smooth continuous control of a more homogeneous surface texture.
Mixed fillers add a further design variable: blends of large magnetically hard and smaller magnetizable particles reach magnetorheological and remanent responses that neither pure filler attains~\cite{moreno-hybrid-MAE}, and whether such a blend also interpolates between the two surface profiles identified here --- with the mixing ratio and the particle size contrast setting where it lands --- is a question the present model is equipped to address.

The model has a well-defined domain of validity.
Below $\xi\approx1.5\times10^{9}$ it systematically over-magnetizes the particles, because the Langevin response of each particle neglects the intra-particle domain interactions and demagnetizing fields that suppress the effective low-field susceptibility of micron-sized multi-domain carbonyl iron particles.
Comparisons with experiment in this low-field regime should be read qualitatively.
The simulated layers are also much thinner than the measured films ($10\sigma$ against roughly $500\sigma$) and the analyzed surface area is some 28 times smaller, so absolute roughness values should be compared in order of magnitude rather than point by point.
Capturing the suppressed low-field susceptibility --- for example through a field-dependent effective susceptibility or an explicit multi-domain particle description --- is the clearest route to quantitative agreement across the full field range.

\section{Methods}\label{sec:methods}

\subsection{Coarse-grained elastomer model}\label{sec:methods-model}

We model the MAE as a dense distribution of magnetic particles interconnected by springs~\cite{pedro-elastomers-surface} (Fig.~\ref{fig: model-sketch}).
This lets us resolve the magnetic interactions at the particle scale while coarse-graining the elastic matrix into a network of springs --- orders of magnitude cheaper than a continuum description.
All molecular dynamics (MD) simulations were carried out using ESPResSo~\cite{espresso-5,espresso-5-zenodo}.

The magnetic particles are modeled as volume-excluding spheres interacting through a Weeks--Chandler--Andersen (WCA) potential energy, Eq.~\eqref{eq:wca}, with parameters $\sigma = 1$ and $\epsilon_{sc} = 1$,
\begin{equation}\label{eq:wca}
    U_{\text{WCA},ij}(r_{ij}) = 4\epsilon_{sc}\!\left[ \left( \frac{\sigma}{r_{ij}} \right)^{\!12} \!\!- \left( \frac{\sigma}{r_{ij}} \right)^{\!6} \!+ \frac{1}{4} \right]
\end{equation}
for $r_{ij}<2^{1/6}\sigma$, and $U_{\text{WCA},ij}=0$ for $r_{ij}\ge2^{1/6}\sigma$.
Here $\sigma$ is the effective particle diameter, $\epsilon_{sc}$ defines the interaction strength --- related to the mechanical softness of the particle --- and $r_{ij}$ is the distance between the particle centers of mass.

Each particle is on average connected by $n_{b} = 6$ harmonic springs,
\begin{align}\label{eq:spr}
    U_{K_{*},ij}(r_{ij})= \dfrac{K_{*,ij}}{2}(r_{ij}-r_{0,ij})^{2},
\end{align}
whose individual spring constants $K_{*}$ are sampled from a truncated and shifted normal distribution within the interval $K_{*} = K_\text{low} = [0.001, 0.01]$ or $K_{*} = K_\text{high} = [0.01, 0.1]$, with mean $\mu_K = \tfrac{1}{2}\!\left(\max(K_{*}) + \min(K_{*})\right)$ and standard deviation $\sigma_K = \tfrac{1}{6}\!\left(\max(K_{*}) - \min(K_{*})\right)$. Only particles closer than the maximum bond equilibrium distance $r_{\text{cutoff}} = 5\sigma$ are considered for spring bonds.
The elastic-matrix properties were tuned through the interval of spring constants $K_{*}$.
For further details see Ref.~\cite{pedro-elastomers-surface}.

\subsection{Magnetically hard and magnetically soft particle models}\label{sec:methods-magnetic}

The magnetically hard particles (MHP) are modeled as permanent point dipoles that relax solely via the physical Brownian rotation of the particle, Eq.~\eqref{eq:mha}.
The magnetically soft particles (MSP) are modeled as magnetizable point dipoles whose magnetic moment is determined by the local magnetic flux density experienced by each particle, $\mathbf{B}_\text{local} = \mathbf{B} + \mathbf{B}_\text{dipolar}$, where $\mathbf{B}$ is the uniform external magnetic flux density and $\mathbf{B}_\text{dipolar}$ is the sum of the fields created by each individual dipole. Here $\mathbf{B}_\text{dipolar}$ is evaluated at the particle center and taken to be locally uniform --- an accurate approximation for particles smaller than $\qty{15}{\nano\meter}$, and still a good approximation for particularly magnetically soft materials such as carbonyl iron regardless of size --- and the particle magnetic moment is then computed using the Langevin function as per Mostarac \textit{et al.}~\cite{deniz-magnetizable-model}, Eq.~\eqref{eq:mso}:
\begin{align}
    \text{MH:} \quad \lVert\boldsymbol{\mu}_{i}\rVert &=\mu_{\infty}, \label{eq:mha} \\
    \notag \\
    \text{MS:} \quad\quad \boldsymbol{\mu}_{i} &=\mu_{\infty}L(\alpha)\,\hat{\mathbf{e}}_{B_{\text{local}}}, \label{eq:mso} \\
    L(\alpha) &=\coth(\alpha) - \frac{1}{\alpha}, \notag \\
    \alpha &=\frac{3\chi_{0}V_{\sigma}}{\mu_{0}\mu_{\infty}}B_{\text{local}}, \notag
\end{align}
where $\mu_{0}$ is the magnetic permeability of vacuum, $\mu_{\infty}$ is the saturation magnetic moment, $\chi_{0}$ is the initial magnetic susceptibility (in the limit $B_{\text{local}}\rightarrow0$), $V_{\sigma}$ is the volume of a single particle, and $B_{\text{local}}\equiv\lVert\mathbf{B}_{\text{local}}\rVert$.

To model particle magnetic interactions, we use the energy of two magnetic dipoles~\cite{Griffiths2023} and the Zeeman energy, understood as the potential energy of a magnetized body in an external magnetic field:
\begin{align}
    U_{\text{dd},ij}(\boldsymbol{r}_{ij},\boldsymbol{\mu}_{i},\boldsymbol{\mu}_{j}) &= \notag \\
    &\hspace{-10mm}=\dfrac{\boldsymbol{\mu}_{i}\cdot\boldsymbol{\mu}_j}{r_{ij}^{3}} - \dfrac{3\left[\boldsymbol{\mu}_{i}\cdot\boldsymbol{r}_{ij}\right] \left[\boldsymbol{\mu}_{j}\cdot\boldsymbol{r}_{ij}\right]}{r_{ij}^{5}}, \label{eq: U_dip}
    \\[1ex]
    U_{\text{Zee},i}(\boldsymbol{\mu}_{i},\boldsymbol{B})&= -\boldsymbol{\mu}_{i}\cdot\boldsymbol{B}, \label{eq: U_Zee}
\end{align}
where $\mathbf{r}_{ij}$ is the vector connecting the centers of particles $i$ and $j$, $r_{ij} \equiv \lVert \mathbf{r}_{ij} \rVert$, and $\boldsymbol{\mu}_i$ is the dipole moment of particle $i$.

\subsection{Sample preparation and simulation protocol}\label{sec:methods-protocol}

To prepare the initial elastomer sample, we first place a substrate of volume-excluding particles at the bottom of the simulation box. The box is then randomly filled with non-overlapping spheres up to the desired height, corresponding to an MAE layer thickness of $10\sigma$.
During this stage, the particles are confined between two walls separated by the target layer thickness and equilibrated for $1 \times 10^{6}$ integration steps using a Langevin thermostat with $k_{\text{B}}\text{T}_{\mathrm{mix}} = 1 \times 10^{-3}$ and $\gamma = 10$. The simulation box sides are $34.73$, $34.73$ and $138.92$ in units of the particle size $\sigma$, for all volume fractions. Each integration step corresponds to a timestep of $0.001$.

After equilibration, the bottommost layer of particles, defined by $z < z_{\mathrm{substrate}} + R_{\text{substrate}} + \tfrac{3}{4}\sigma$, where $R_{\text{substrate}}$ is the radius of the substrate particles, is adhered to the substrate by constraining their motion to the $xy$-plane while keeping their current $z$ coordinate fixed. The elastomer is then cured by randomly connecting particles with springs, as described in the \textit{Methods} (``Coarse-grained elastomer model''), and the chosen magnetic model is assigned to each particle.

For MAEs containing MHP, an additional equilibration of $5 \times 10^{4}$ integration steps relaxes the dipolar interactions.
This step is unnecessary for MSP, since they are unmagnetized in the absence of an external field and are therefore already in equilibrium. For all particles we set $\mu_{\infty}=1$, while MS particles are assigned $\chi_0 = 2.4$.
The thermostat is then changed to $k_{\text{B}}\text{T}_{\mathrm{sim}} = 1 \times 10^{-6}$ and $\gamma = 100$.
This low temperature is chosen because, in the experiment, the contribution of entropy is negligible, while for computational efficiency the Langevin equations are still solved; this turns out to be a suitable approach for modeling MAEs~\cite{pedro-elastomers-surface,dobroserdova2026influence}.

Starting from this initial MAE configuration, we apply a uniform magnetic field $\mathbf{B} = B\mathbf{e}_{z}$ and allow the system to relax for $2.5 \times 10^{5}$ integration steps.
All reported field values correspond to independent runs started from the $B=0$ configuration, so no field history is accumulated.

\subsection{Characteristic energies}\label{sec:methods-energies}

The interactions in the system are governed by three potential energies: harmonic spring (Eq.~\eqref{eq:spr}), dipolar (Eq.~\eqref{eq: U_dip}) and Zeeman (Eq.~\eqref{eq: U_Zee}).
To compare between systems with different numbers of particles, we report per-particle averages, with $N$ the total number of particles in each simulation.
All energies are later non-dimensionalized by dividing by the thermal energy $k_{\text{B}}\text{T}$.

The per-particle averaged elastic (bond) energy is
\begin{equation}
    \langle E_{\text{bond}}\rangle = \dfrac{1}{N} \sum_{\langle i,j\rangle} U_{K,ij}, \label{eq: E_bond}
\end{equation}
where $\langle i,j\rangle$ runs over all bonded particle pairs, and $U_{K,ij}$ is the elastic potential energy between particles $i$ and $j$ (defined in Eq.~\eqref{eq:spr}).

The per-particle averaged Zeeman energy is
\begin{equation}
    \langle E_{\text{Zeeman}}\rangle = \dfrac{1}{N} \sum_{i} U_{\text{Zee},i}, \label{eq: E_Zee}
\end{equation}
where $i$ runs over all particles, and $U_{\text{Zee},i}$ is the Zeeman potential energy of particle $i$ (defined in Eq.~\eqref{eq: U_Zee}).
The main information carried by the Zeeman energy is the alignment of the particles with the applied field.
To directly compare the alignment of MHP and MSP with the applied field, we define the per-particle averaged $\mu$-normalized Zeeman energy as
\begin{equation}
    \langle E_{\text{Zeeman},\mu}\rangle = -\dfrac{1}{N} \sum_{i} \dfrac{\boldsymbol{\mu}_{i}}{\mu_{i}}\cdot\mathbf{B}, \label{eq: E_Zee_mu}
\end{equation}
with $\mu_{i}\equiv \lVert \boldsymbol{\mu}_{i} \rVert$.
For a dipole perfectly aligned with $\mathbf{B}$, $\langle E_{\text{Zeeman},\mu}\rangle = -\lVert \mathbf{B}\rVert$.

The per-particle averaged dipolar energy is
\begin{equation}
    \langle E_{\text{dipolar}}\rangle = \dfrac{1}{N} \sum_{i<j} U_{\text{dd},ij}, \label{eq: E_dip}
\end{equation}
where $i<j$ runs over all unique pairs of particles, and $U_{\text{dd},ij}$ is the dipole--dipole potential energy between particles $i$ and $j$ (defined in Eq.~\eqref{eq: U_dip}).

\subsection{Magnetic moment and magnetization}\label{sec:methods-magnetization}

To characterize the magnetic response of the layer, we compute the average $z$-component of the magnetic moment per particle, normalized by the saturation moment,
\begin{equation}
    \langle \mu_{z}\rangle/\mu_{\infty} = \dfrac{1}{N\mu_{\infty}} \sum_{i} \boldsymbol{\mu}_{i}\cdot \hat{\mathbf{e}}_{z},
\end{equation}
which measures how close the ensemble is to full magnetization along the field direction.

We also compute the magnetization of the layer, normalized by the bulk saturation magnetization of the particle material,
\begin{equation}
    M/M_{\text{s}} = \varphi \langle \mu_{z}\rangle/\mu_{\infty},
\end{equation}
where $\varphi$ is the volume fraction of magnetic particles and $M_{\text{s}}=\mu_{\infty}/V_{\sigma}$ is the bulk material saturation magnetization.

Finally, to separate the degree of magnetization from the degree of alignment, we compute
\begin{equation}
    \langle \mu_{z}/\mu\rangle = \dfrac{1}{N} \sum_{i} \dfrac{\boldsymbol{\mu}_{i}}{\mu_{i}}\cdot\hat{\mathbf{e}}_{z},
\end{equation}
which equals $1$ for perfectly field-aligned moments regardless of their magnitude.

\subsection{Surface reconstruction and roughness}\label{sec:methods-roughness}

To characterize the roughness of the MAE layer surface, we reconstruct the surface using the Identification of Truly Interfacial Molecules (ITIM) approach~\cite{ITIM-original} (Figs.~\ref{fig: surface-roughness}~b,~c) and then compute the root-mean-square roughness parameter,
\begin{equation}
    R_{\text{rms}} = \sqrt{
        \dfrac{1}{N_{X}N_{Y}} \sum_{i, j} \left[ \text{height}(x_{i}, y_{j}) - \left< \text{height} \right> \right]^{2}}. \label{eq: R_rms}
\end{equation}
Here $N_{X}$ and $N_{Y}$ are the number of ITIM grid points in the $x$- and $y$-directions, respectively, $i$ and $j$ run over the grid point coordinates, and $\text{height}(x_{i},y_{j})$ is the height of the surface at position $(x_{i}, y_{j})$.
The roughness parameter is later non-dimensionalized by the particle size $\sigma$.

\subsection{Bond-order parameters}\label{sec:methods-bop}

The local Steinhardt order parameters are computed as
\begin{equation}
q_{\ell m}(i) = \frac{1}{N_b(i)} \sum_{j=1}^{N_b(i)} Y_{\ell m}(\hat{\mathbf{r}}_{ij}),
\end{equation}
where $N_b(i)$ is the number of neighbors of particle $i$, $Y_{\ell m}$ are spherical harmonics, and $\hat{\mathbf{r}}_{ij}$ is the unit vector connecting particles $i$ and $j$~\cite{steinhardt83bond}.
Two particles are considered neighbors if their center-to-center distance is smaller than $1.3\sigma$.
The rotationally invariant scalar bond-order parameter is
\begin{equation}
q_{\ell}(i) = \sqrt{\frac{4\pi}{2\ell+1} \sum_{m=-\ell}^{\ell} \left| q_{\ell m}(i) \right|^2 }.
\end{equation}

To reduce noise and improve robustness, we use the Lechner--Dellago averaged version~\cite{lechner08accurate}, in which the spherical harmonic components are averaged over the first neighbor shell,
\begin{equation}
\bar{q}_{\ell m}(i) = \frac{1}{N_b(i)+1} \sum_{k \in \{i \cup N(i)\}} q_{\ell m}(k),
\end{equation}
with the corresponding averaged bond-order parameter
\begin{equation}
\bar{q}_{\ell}(i) = \sqrt{\frac{4\pi}{2\ell+1} \sum_{m=-\ell}^{\ell} \left| \bar{q}_{\ell m}(i) \right|^2 }. \label{eq: BOP}
\end{equation}

We calculate $\bar{q}_{\ell}(i)$ only for particles inside of the bulk of the MAE slab.
To separate the bulk particles, we compute the particle density profile along $z$,
\begin{equation}
    \rho_{z}(z_{k}) = \dfrac{N_{k}}{A\Delta z},
\end{equation}
binning the layer along $z$ and dividing the number of particles in bin $k$ --- those between $z=(k-1)\Delta z$ and $z=k\Delta z$ --- by the bin volume $A\Delta z$, and select the region where this quantity is approximately constant.

The average value of the averaged bond-order parameters for the bulk particles is
\begin{equation}
    \langle \bar{q}_{\ell} \rangle = \dfrac{1}{N_{\text{bulk}}} \sum_{i} \bar{q}_{\ell}(i), \label{eq: BOP_avg}
\end{equation}
where $N_{\text{bulk}}$ is the number of bulk particles and here $i$ runs only over the bulk particles.

\subsection{Mapping to physical units}\label{sec:methods-units}

We use non-dimensional quantities to compare measurements across different unit systems, taking the average particle diameter $\sigma$, the saturation magnetic moment $\mu_{\infty}$, the bulk saturation magnetization $M_{\text{s}}$ and the thermal energy $k_{\text{B}}\text{T}$ as base scales for length, magnetic moment, magnetization and energy.
The non-dimensional field is the Langevin parameter $\xi=\mu_{\infty}B/k_{\text{B}}\text{T}$.

The simulation temperature is already close to zero, $k_{\text{B}}\text{T}_{\text{sim}}=10^{-6}$, so the system is essentially athermal.
We exploit this by treating $k_{\text{B}}\text{T}$ as a free parameter that scales the non-dimensional field onto the experimental field axis of the carbonyl iron particles.
We fix it from a single anchor point: we take the average magnetic moment at the largest simulated field, find the experimental field at which the measured moment matches that value, and solve for the $k_{\text{B}}\text{T}$ that places the two points at the same physical field. For example, if the largest simulated field gives an average moment of $0.85$, we locate the experimental point for the moment closest to $0.85$ and compute the $k_{\text{B}}\text{T}$ that best aligns the two field axes (Fig.~\ref{fig: mag_exp}).
The \emph{shape} of the magnetization curve is set entirely by the particle susceptibility $\chi_{0}$, diameter $\sigma$ and volume fraction $\varphi$ through the Langevin response and the dipolar fields (Eq.~\eqref{eq:mso}) and is not adjusted in this procedure; only the field unit is matched.
Since this is a single horizontal rescaling fixed by one point rather than a per-curve fit, agreement at any other field constitutes a genuine test of the model rather than a consequence of fitting.
A single value of $k_{\text{B}}\text{T}\approx1.61\times10^{-8}$ aligns the curves consistently across all measured volume fractions; the rescaling is not tuned per density.
We stress that the mismatch between the simulated temperature $k_{\text{B}}\text{T}=10^{-6}$ and this fitted value has no consequence for the particle configurations, since in both cases the system experiences no effects of thermal noise.
The resulting conversion between $\xi$, SI and simulation units is given in Supplementary Table~\ref{tab: xi-to-B}.

\subsection{Experimental sample preparation}\label{sec:methods-exp-samples}

Isotropic MAEs were prepared using a procedure adapted from methodologies reported in earlier studies~\cite{kriegl2022microstructured,belyaeva2017magnetodielectric}.
In the first step, the base polymer VS 100000 was blended with MV 2000, Modifier 715, and the silicone oil WACKER\textsuperscript{\textregistered} AK 10 using a hand-held electric mixer (Roti\textsuperscript{\textregistered}-Speed-stirrer, Carl Roth GmbH, Germany) operated at approximately \qty{5000}{rpm} for $\qty{2}{\minute}$, producing a homogeneous base compound.
The carbonyl iron filler was then incorporated into this compound using a planetary centrifugal mixer (Thinky\textsuperscript{\textregistered} ARV-501, Thinky Corporation, Japan) at \qty{800}{rpm} for \qty{60}{\second} to ensure uniform dispersion. Subsequently, Crosslinker 210, Pt-catalyst 510, and inhibitor DVS were added, and the mixture was processed under vacuum in the same planetary mixer at \qty{800}{rpm} for \qty{90}{\second}. The resulting compound was deposited on a thin PET film ($\approx\qty{0.1}{\milli\meter}$ thickness) and cast using a doctor-blade applicator to ensure uniform film thickness, yielding films with a final thickness of \qty{2}{\milli\meter}.
Curing was carried out in an air-circulating oven at \qty{80}{\degreeCelsius} for \qty{1}{\hour} and then at \qty{65}{\degreeCelsius} for \qty{24}{\hour}. After curing, films were covered with an additional PET sheet to prevent surface contamination prior to further characterization.

\subsection{Experimental surface characterization}\label{sec:methods-exp-surface}

Experimental surface roughnesses were calculated from surface topography measurements using a confocal laser scanning microscope (LEXT OLS4000, Olympus, Germany) equipped with a $20\times$ objective. Large-area surface maps of \qty{1.8}{\milli\meter}\,$\times$\,\qty{1.2}{\milli\meter} (roughly $450\times300$ average particle diameters) were obtained through image stitching and subsequently processed using polynomial background subtraction and median filtering. Surface roughness was evaluated under different magnetic flux densities and filler concentrations.
For comparison, the simulated surface roughnesses were calculated from the four ITIM-estimated surfaces (as in Fig.~\ref{fig: surface-roughness}~c), which together span roughly $70\times70$ particle diameters --- an area about $28$ times smaller than the experimental surface area --- and the experimental films are much thicker (around $500\sigma$) than the simulated layer ($10\sigma$).

\section*{Data availability}\label{sec:data}


The simulation data supporting the findings of this study --- the initial and final particle configurations of every simulated system together with the derived observables --- are available at~\cite{paper1-data-zenodo}.
The experimental magnetization curves used for validation in Fig.~\ref{fig: mag_exp} are published data, taken from Refs.~\cite{gasper-magnetisation,gasper-magnetisation-zenodo}.
The experimental surface roughness and stiffness data are deposited at Mendeley Data~\cite{mahdi-data-E-and-R}.

\section*{Code availability}\label{sec:code}


The molecular dynamics simulations were performed with the open-source package ESPResSo~\cite{espresso-5,espresso-5-zenodo} and the open-source package Pressomancy~\cite{pressomancy-github}, which leverages ESPResSo routines.
The exact custom ESPResSo and pressomancy codebase, the custom simulation and analysis scripts implementing the magnetizable-particle MAE model, the ITIM surface reconstruction workflow and the bond-order-parameter analysis are available at~\cite{paper1-code-repo}, together with the ESPResSo source tree used.



\begin{thebibliography}{10}
\expandafter\ifx\csname url\endcsname\relax
  \def\url#1{\burl{#1}}\fi
\expandafter\ifx\csname urlprefix\endcsname\relax\def\urlprefix{URL }\fi
\providecommand{\bibinfo}[2]{#2}
\providecommand{\eprint}[2][]{\url{#2}}
\providecommand{\doi}[1]{\url{https://doi.org/#1}}
\bibcommenthead

\bibitem{review-magnetic-soft-matter-and-robots}
\bibinfo{author}{Kim, Y.} \& \bibinfo{author}{Zhao, X.}
\newblock \bibinfo{title}{Magnetic soft materials and robots}.
\newblock \emph{\bibinfo{journal}{Chemical Reviews}}
  \textbf{\bibinfo{volume}{122}}, \bibinfo{pages}{5317--5364}
  (\bibinfo{year}{2022}).
\newblock \urlprefix\url{https://doi.org/10.1021/acs.chemrev.1c00481}.
\newblock \bibinfo{note}{PMID: 35104403}.

\bibitem{soft-robotic-review-2}
\bibinfo{author}{Yasa, O.} \emph{et~al.}
\newblock \bibinfo{title}{An overview of soft robotics}.
\newblock \emph{\bibinfo{journal}{Annual Review of Control, Robotics, and
  Autonomous Systems}} \textbf{\bibinfo{volume}{6}}, \bibinfo{pages}{1--29}
  (\bibinfo{year}{2023}).
\newblock
  \urlprefix\url{https://www.annualreviews.org/content/journals/10.1146/annurev-control-062322-100607}.

\bibitem{magnetic-soft-polymer-composites}
\bibinfo{author}{Filipcsei, G.}, \bibinfo{author}{Csetneki, I.},
  \bibinfo{author}{Szil{\'a}gyi, A.} \& \bibinfo{author}{Zr{\'i}nyi, M.}
\newblock \emph{\bibinfo{title}{Magnetic Field-Responsive Smart Polymer
  Composites}}, \bibinfo{pages}{137--189} (\bibinfo{publisher}{Springer Berlin
  Heidelberg}, \bibinfo{address}{Berlin, Heidelberg}, \bibinfo{year}{2007}).
\newblock \urlprefix\url{https://doi.org/10.1007/12_2006_104}.

\bibitem{1995-shiga}
\bibinfo{author}{Shiga, T.}, \bibinfo{author}{Okada, A.} \&
  \bibinfo{author}{Kurauchi, T.}
\newblock \bibinfo{title}{Magnetroviscoelastic behavior of composite gels}.
\newblock \emph{\bibinfo{journal}{J. Appl. Polym. Sci.}}
  \textbf{\bibinfo{volume}{58}}, \bibinfo{pages}{787--792}
  (\bibinfo{year}{1995}).

\bibitem{2000-zrinyi}
\bibinfo{author}{Zr\'inyi, M.}
\newblock \bibinfo{title}{Intelligent polymer gels controlled by magnetic
  fields}.
\newblock \emph{\bibinfo{journal}{Colloid Polym. Sci.}}
  \textbf{\bibinfo{volume}{278}}, \bibinfo{pages}{98--103}
  (\bibinfo{year}{2000}).

\bibitem{Stepanov2008}
\bibinfo{author}{Stepanov, G.}, \bibinfo{author}{Borin, D.},
  \bibinfo{author}{Raikher, Y.}, \bibinfo{author}{Melenev, P.} \&
  \bibinfo{author}{Perov, N.}
\newblock \bibinfo{title}{Motion of ferroparticles inside the polymeric matrix
  in magnetoactive elastomers}.
\newblock \emph{\bibinfo{journal}{Journal of Physics: Condensed Matter}}
  \textbf{\bibinfo{volume}{20}}, \bibinfo{pages}{204121}
  (\bibinfo{year}{2008}).

\bibitem{Wereley2014}
\bibinfo{editor}{Wereley, N.} (ed.) \emph{\bibinfo{title}{Magnetorheology:
  advances and applications}}  (\bibinfo{publisher}{RSC Publishing},
  \bibinfo{address}{Cambridge}, \bibinfo{year}{2014}).

\bibitem{Morillas2020}
\bibinfo{author}{Morillas, J.~R.} \& \bibinfo{author}{de~Vicente, J.}
\newblock \bibinfo{title}{Magnetorheology: a review}.
\newblock \emph{\bibinfo{journal}{Soft Matter}} \textbf{\bibinfo{volume}{16}},
  \bibinfo{pages}{9614--9642} (\bibinfo{year}{2020}).
\newblock \urlprefix\url{http://dx.doi.org/10.1039/D0SM01082K}.

\bibitem{1996-jolly}
\bibinfo{author}{Jolly, M.~R.}, \bibinfo{author}{Carlson, J.~D.},
  \bibinfo{author}{Muñoz, B.~C.} \& \bibinfo{author}{Bullions, T.~A.}
\newblock \bibinfo{title}{The magnetoviscoelastic response of elastomer
  composites consisting of ferrous particles embedded in a polymer matrix}.
\newblock \emph{\bibinfo{journal}{J. Intell. Mater. Syst. Struct.}}
  \textbf{\bibinfo{volume}{7}}, \bibinfo{pages}{613--622}
  (\bibinfo{year}{1996}).

\bibitem{1999-ginder}
\bibinfo{author}{Ginder, J.~M.}, \bibinfo{author}{Nichols, M.~E.},
  \bibinfo{author}{Elie, L.~D.} \& \bibinfo{author}{Tardiff, J.~L.}
\newblock \bibinfo{editor}{Wuttig, M.~R.} (ed.)
  \emph{\bibinfo{title}{Magnetorheological elastomers: properties and
  applications}}.
\newblock (ed.\bibinfo{editor}{Wuttig, M.~R.}) \emph{\bibinfo{booktitle}{Smart
  Structures and Materials 1999: Smart Materials Technologies}}, Vol.
  \bibinfo{volume}{3675} of \emph{\bibinfo{series}{Proceedings SPIE}},
  \bibinfo{pages}{131--138} (\bibinfo{publisher}{SPIE}, \bibinfo{year}{1999}).

\bibitem{2000-carlson}
\bibinfo{author}{Carlson, J.} \& \bibinfo{author}{Jolly, M.~R.}
\newblock \bibinfo{title}{Mr fluid, foam and elastomer devices}.
\newblock \emph{\bibinfo{journal}{Mechatronics}} \textbf{\bibinfo{volume}{10}},
  \bibinfo{pages}{555--569} (\bibinfo{year}{2000}).

\bibitem{2006-varga}
\bibinfo{author}{Varga, Z.}, \bibinfo{author}{Filipcsei, G.} \&
  \bibinfo{author}{Zr\'inyi, M.}
\newblock \bibinfo{title}{Magnetic field sensitive functional elastomers with
  tuneable elastic modulus}.
\newblock \emph{\bibinfo{journal}{Polymer}} \textbf{\bibinfo{volume}{47}},
  \bibinfo{pages}{227--233} (\bibinfo{year}{2006}).

\bibitem{2012-boczkowska}
\bibinfo{author}{Boczkowska, A.} \& \bibinfo{author}{Awietjan, S.}
\newblock \bibinfo{title}{ in \textit{Microstructure and properties of
  magnetorheological elastomers}} (ed.\bibinfo{editor}{Boczkowska, A.})
  \emph{\bibinfo{booktitle}{Advanced Elastomers - Technology, Properties and
  Applications}} Ch.~\bibinfo{chapter}{06}, \bibinfo{pages}{414}
  (\bibinfo{publisher}{InTech}, \bibinfo{address}{Rijeka},
  \bibinfo{year}{2012}).

\bibitem{2016-odenbach}
\bibinfo{author}{Odenbach, S.}
\newblock \bibinfo{title}{Microstructure and rheology of magnetic hybrid
  materials}.
\newblock \emph{\bibinfo{journal}{Arch. Appl. Mech.}}
  \textbf{\bibinfo{volume}{86}}, \bibinfo{pages}{269--279}
  (\bibinfo{year}{2016}).

\bibitem{2006-deng}
\bibinfo{author}{Deng, H.-X.}, \bibinfo{author}{Gong, X.-L.} \&
  \bibinfo{author}{Wang, L.-H.}
\newblock \bibinfo{title}{Development of an adaptive tuned vibration absorber
  with magnetorheological elastomer}.
\newblock \emph{\bibinfo{journal}{Smart Mater. Struct.}}
  \textbf{\bibinfo{volume}{15}}, \bibinfo{pages}{N111} (\bibinfo{year}{2006}).

\bibitem{2013-li}
\bibinfo{author}{Li, W.~H.}, \bibinfo{author}{Zhang, X.~Z.} \&
  \bibinfo{author}{Du, H.}
\newblock \emph{\bibinfo{title}{Magnetorheological Elastomers and Their
  Applications}}, Vol.~\bibinfo{volume}{11} of \emph{\bibinfo{series}{Advanced
  Structured Materials}}, \bibinfo{pages}{357--374}
  (\bibinfo{publisher}{Springer Berlin Heidelberg}, \bibinfo{address}{Berlin,
  Heidelberg}, \bibinfo{year}{2013}).

\bibitem{2014-li}
\bibinfo{author}{Li, Y.}, \bibinfo{author}{Li, J.}, \bibinfo{author}{Li, W.} \&
  \bibinfo{author}{Du, H.}
\newblock \bibinfo{title}{A state-of-the-art review on magnetorheological
  elastomer devices}.
\newblock \emph{\bibinfo{journal}{Smart Mater. Struct.}}
  \textbf{\bibinfo{volume}{23}}, \bibinfo{pages}{123001}
  (\bibinfo{year}{2014}).

\bibitem{shamonin-stiffness}
\bibinfo{author}{Stoll, A.}, \bibinfo{author}{Mayer, M.},
  \bibinfo{author}{Monkman, G.~J.} \& \bibinfo{author}{Shamonin, M.}
\newblock \bibinfo{title}{Evaluation of highly compliant magneto-active
  elastomers with colossal magnetorheological response}.
\newblock \emph{\bibinfo{journal}{Journal of Applied Polymer Science}}
  \textbf{\bibinfo{volume}{131}} (\bibinfo{year}{2014}).
\newblock
  \urlprefix\url{https://onlinelibrary.wiley.com/doi/abs/10.1002/app.39793}.

\bibitem{roughness-gasper}
\bibinfo{author}{Glavan, G.}, \bibinfo{author}{Kettl, W.},
  \bibinfo{author}{Brunhuber, A.}, \bibinfo{author}{Shamonin, M.} \&
  \bibinfo{author}{Drevenšek-Olenik, I.}
\newblock \bibinfo{title}{Effect of material composition on tunable surface
  roughness of magnetoactive elastomers}.
\newblock \emph{\bibinfo{journal}{Polymers}} \textbf{\bibinfo{volume}{11}}
  (\bibinfo{year}{2019}).
\newblock \urlprefix\url{https://www.mdpi.com/2073-4360/11/4/594}.

\bibitem{surface-roughness-kramarenko}
\bibinfo{author}{Kirgizov, S.~E.}, \bibinfo{author}{Kostrov, S.~A.} \&
  \bibinfo{author}{Kramarenko, E.~Y.}
\newblock \bibinfo{title}{Magneto-tunable surface roughness and hydrophobicity
  of magnetoactive elastomers based on polymer networks with different
  architectures}.
\newblock \emph{\bibinfo{journal}{Polymers}} \textbf{\bibinfo{volume}{17}}
  (\bibinfo{year}{2025}).
\newblock \urlprefix\url{https://www.mdpi.com/2073-4360/17/17/2411}.

\bibitem{li-roughness-increase-and-decrease}
\bibinfo{author}{Li, R.} \emph{et~al.}
\newblock \bibinfo{title}{Study on sliding friction characteristics of
  magnetorheological elastomer—copper pair affected by magnetic-controlled
  surface roughness and elastic modulus}.
\newblock \emph{\bibinfo{journal}{Smart Materials and Structures}}
  \textbf{\bibinfo{volume}{31}}, \bibinfo{pages}{015030}
  (\bibinfo{year}{2021}).
\newblock \urlprefix\url{https://doi.org/10.1088/1361-665X/ac3c05}.

\bibitem{drotlefMagneticallyActuatedPatterns2014}
\bibinfo{author}{Drotlef, D.-M.}, \bibinfo{author}{Bl{\"u}mler, P.} \&
  \bibinfo{author}{Del~Campo, A.}
\newblock \bibinfo{title}{Magnetically actuated patterns for bioinspired
  reversible adhesion (dry and wet)}.
\newblock \emph{\bibinfo{journal}{Advanced materials (Deerfield Beach, Fla.)}}
  \textbf{\bibinfo{volume}{26}}, \bibinfo{pages}{775--9}
  (\bibinfo{year}{2014}).

\bibitem{kovalevMagneticallySwitchableAdhesion2022}
\bibinfo{author}{Kovalev, A.}, \bibinfo{author}{Belyaeva, I.~A.},
  \bibinfo{author}{{von Hofen}, C.}, \bibinfo{author}{Gorb, S.} \&
  \bibinfo{author}{Shamonin, M.}
\newblock \bibinfo{title}{Magnetically {{Switchable Adhesion}} and {{Friction}}
  of {{Soft Magnetoactive Elastomers}}}.
\newblock \emph{\bibinfo{journal}{Advanced Engineering Materials}}
  \bibinfo{pages}{2200372} (\bibinfo{year}{2022}).

\bibitem{krieglTunableContactAngle2023}
\bibinfo{author}{Kriegl, R.}, \bibinfo{author}{Kovalev, A.},
  \bibinfo{author}{Shamonin, M.} \& \bibinfo{author}{Gorb, S.}
\newblock \bibinfo{title}{Tunable contact angle hysteresis on compliant
  magnetoactive elastomers}.
\newblock \emph{\bibinfo{journal}{Extreme Mechanics Letters}}
  \textbf{\bibinfo{volume}{63}}, \bibinfo{pages}{102049}
  (\bibinfo{year}{2023}).

\bibitem{moreno-hybrid-MAE}
\bibinfo{author}{Moreno-Mateos, M.~A.}, \bibinfo{author}{Hossain, M.},
  \bibinfo{author}{Steinmann, P.} \& \bibinfo{author}{Garcia-Gonzalez, D.}
\newblock \bibinfo{title}{Hybrid magnetorheological elastomers enable versatile
  soft actuators}.
\newblock \emph{\bibinfo{journal}{npj Computational Materials}}
  \textbf{\bibinfo{volume}{8}}, \bibinfo{pages}{162} (\bibinfo{year}{2022}).
\newblock \urlprefix\url{https://doi.org/10.1038/s41524-022-00844-1}.

\bibitem{2008-guang}
\bibinfo{author}{Guan, X.}, \bibinfo{author}{Dong, X.} \& \bibinfo{author}{Ou,
  J.}
\newblock \bibinfo{title}{Magnetostrictive effect of magnetorheological
  elastomer}.
\newblock \emph{\bibinfo{journal}{J. Magn. Magn. Mater.}}
  \textbf{\bibinfo{volume}{320}}, \bibinfo{pages}{158--163}
  (\bibinfo{year}{2008}).

\bibitem{Diguet2010}
\bibinfo{author}{Diguet, G.}, \bibinfo{author}{Beaugnon, E.} \&
  \bibinfo{author}{Cavaillé, J.~Y.}
\newblock \bibinfo{title}{Shape effect in the magnetostriction of ferromagnetic
  composite}.
\newblock \emph{\bibinfo{journal}{J. Magn. Magn. Mater.}}
  \textbf{\bibinfo{volume}{322}}, \bibinfo{pages}{3337--3341}
  (\bibinfo{year}{2010}).

\bibitem{silva2022giant}
\bibinfo{author}{Silva, J.}, \bibinfo{author}{Gouveia, C.},
  \bibinfo{author}{Dinis, G.}, \bibinfo{author}{Pinto, A.} \&
  \bibinfo{author}{Pereira, A.}
\newblock \bibinfo{title}{Giant magnetostriction in low-concentration
  magnetorheological elastomers}.
\newblock \emph{\bibinfo{journal}{Composites Part B: Engineering}}
  \textbf{\bibinfo{volume}{243}}, \bibinfo{pages}{110125}
  (\bibinfo{year}{2022}).
\newblock
  \urlprefix\url{https://www.sciencedirect.com/science/article/pii/S1359836822005017}.

\bibitem{Glavan2024}
\bibinfo{author}{Glavan, G.}, \bibinfo{author}{Belyaeva, I.~A.} \&
  \bibinfo{author}{Shamonin, M.}
\newblock \bibinfo{title}{On the piezomagnetism of magnetoactive elastomeric
  cylinders in uniform magnetic fields: Height modulation in the vicinity of an
  operating point by time-harmonic fields}.
\newblock \emph{\bibinfo{journal}{Polymers}} \textbf{\bibinfo{volume}{16}},
  \bibinfo{pages}{2706} (\bibinfo{year}{2024}).

\bibitem{romeis-model-susceptibility-130}
\bibinfo{author}{Roghani, M.} \emph{et~al.}
\newblock \bibinfo{title}{Magnetically induced deformation of isotropic
  magnetoactive elastomers and its relation to the magnetorheological effect}.
\newblock \emph{\bibinfo{journal}{Phys. Rev. Appl.}}
  \textbf{\bibinfo{volume}{23}}, \bibinfo{pages}{034041}
  (\bibinfo{year}{2025}).
\newblock
  \urlprefix\url{https://link.aps.org/doi/10.1103/PhysRevApplied.23.034041}.

\bibitem{2014-pessot}
\bibinfo{author}{Pessot, G.} \emph{et~al.}
\newblock \bibinfo{title}{Structural control of elastic moduli in ferrogels and
  the importance of non-affine deformations}.
\newblock \emph{\bibinfo{journal}{J. Chem. Phys.}}
  \textbf{\bibinfo{volume}{141}}, \bibinfo{pages}{124904}
  (\bibinfo{year}{2014}).

\bibitem{2016-pessot}
\bibinfo{author}{Pessot, G.}, \bibinfo{author}{L\"owen, H.} \&
  \bibinfo{author}{Menzel, A.~M.}
\newblock \bibinfo{title}{Dynamic elastic moduli in magnetic gels: Normal modes
  and linear response}.
\newblock \emph{\bibinfo{journal}{J. Chem. Phys.}}
  \textbf{\bibinfo{volume}{145}}, \bibinfo{pages}{104904}
  (\bibinfo{year}{2016}).

\bibitem{Snarskii2021}
\bibinfo{author}{Snarskii, A.~A.}, \bibinfo{author}{Shamonin, M.} \&
  \bibinfo{author}{Yuskevich, P.}
\newblock \bibinfo{title}{Effect of magnetic-field-induced restructuring on the
  elastic properties of magnetoactive elastomers}.
\newblock \emph{\bibinfo{journal}{J. Magn. Magn. Mater.}}
  \textbf{\bibinfo{volume}{517}}, \bibinfo{pages}{167392}
  (\bibinfo{year}{2021}).

\bibitem{sorokin2018controllable}
\bibinfo{author}{Sorokin, V.~V.}, \bibinfo{author}{Sokolov, B.~O.},
  \bibinfo{author}{Stepanov, G.~V.} \& \bibinfo{author}{Kramarenko, E.~Y.}
\newblock \bibinfo{title}{Controllable hydrophobicity of magnetoactive
  elastomer coatings}.
\newblock \emph{\bibinfo{journal}{Journal of Magnetism and Magnetic Materials}}
  \textbf{\bibinfo{volume}{459}}, \bibinfo{pages}{268--271}
  (\bibinfo{year}{2018}).
\newblock
  \urlprefix\url{https://www.sciencedirect.com/science/article/pii/S0304885317321546}.
\newblock \bibinfo{note}{The selected papers of Seventh Moscow International
  Symposium on Magnetism (MISM-2017)}.

\bibitem{chen2021magnetic}
\bibinfo{author}{Chen, S.}, \bibinfo{author}{Zhu, M.}, \bibinfo{author}{Zhang,
  Y.}, \bibinfo{author}{Dong, S.} \& \bibinfo{author}{Wang, X.}
\newblock \bibinfo{title}{Magnetic-responsive superhydrophobic surface of
  magnetorheological elastomers mimicking from lotus leaves to rose petals}.
\newblock \emph{\bibinfo{journal}{Langmuir}} \textbf{\bibinfo{volume}{37}},
  \bibinfo{pages}{2312--2321} (\bibinfo{year}{2021}).

\bibitem{zhangControllableMagneticRoughness2020}
\bibinfo{author}{Zhang, H.}, \bibinfo{author}{Tao, Z.} \&
  \bibinfo{author}{Xiao, Y.}
\newblock \bibinfo{title}{Controllable magnetic roughness surface with
  sustainable superhydrophobicity based on magnetorheological colloid}.
\newblock \emph{\bibinfo{journal}{Journal of Intelligent Material Systems and
  Structures}} \bibinfo{pages}{10453892094283} (\bibinfo{year}{2020}).

\bibitem{kriegl2022microstructured}
\bibinfo{author}{Kriegl, R.} \emph{et~al.}
\newblock \bibinfo{title}{Microstructured magnetoactive elastomers for
  switchable wettability}.
\newblock \emph{\bibinfo{journal}{Polymers}} \textbf{\bibinfo{volume}{14}},
  \bibinfo{pages}{3883} (\bibinfo{year}{2022}).

\bibitem{Borbath2012}
\bibinfo{author}{Borbath, T.}, \bibinfo{author}{G{\"u}nther, S.},
  \bibinfo{author}{Borin, D.}, \bibinfo{author}{G{\"u}ndermann, T.} \&
  \bibinfo{author}{Odenbach, S.}
\newblock \bibinfo{title}{$\text{X}\mu\text{CT}$ analysis of magnetic
  field-induced phase transitions in magnetorheological elastomers}.
\newblock \emph{\bibinfo{journal}{Smart Materials and Structures}}
  \textbf{\bibinfo{volume}{21}}, \bibinfo{pages}{105018}
  (\bibinfo{year}{2012}).
\newblock \bibinfo{note}{Cited By 89}.

\bibitem{Schumann201788}
\bibinfo{author}{Sch{\"u}mann, M.} \& \bibinfo{author}{Odenbach, S.}
\newblock \bibinfo{title}{In-situ observation of the particle microstructure of
  magnetorheological elastomers in presence of mechanical strain and magnetic
  fields}.
\newblock \emph{\bibinfo{journal}{Journal of Magnetism and Magnetic Materials}}
  \textbf{\bibinfo{volume}{441}}, \bibinfo{pages}{88--92}
  (\bibinfo{year}{2017}).
\newblock \bibinfo{note}{Cited By 43}.

\bibitem{Flakes_Malte_2017}
\bibinfo{author}{Sch\"umann, M.} \emph{et~al.}
\newblock \bibinfo{title}{A characterisation of the magnetically induced
  movement of ndfeb-particles in magnetorheological elastomers}.
\newblock \emph{\bibinfo{journal}{Smart Materials and Structures}}
  \textbf{\bibinfo{volume}{26}}, \bibinfo{pages}{095018}
  (\bibinfo{year}{2017}).

\bibitem{raikher2008shape}
\bibinfo{author}{Raikher, Y.~L.}, \bibinfo{author}{Stolbov, O.} \&
  \bibinfo{author}{Stepanov, G.}
\newblock \bibinfo{title}{Shape instability of a magnetic elastomer membrane}.
\newblock \emph{\bibinfo{journal}{Journal of Physics D: Applied Physics}}
  \textbf{\bibinfo{volume}{41}}, \bibinfo{pages}{152002}
  (\bibinfo{year}{2008}).

\bibitem{ivaneyko2012effects}
\bibinfo{author}{Ivaneyko, D.}, \bibinfo{author}{Toshchevikov, V.},
  \bibinfo{author}{Saphiannikova, M.} \& \bibinfo{author}{Heinrich, G.}
\newblock \bibinfo{title}{Effects of particle distribution on mechanical
  properties of magneto-sensitive elastomers in a homogeneous magnetic field}.
\newblock \emph{\bibinfo{journal}{arXiv preprint arXiv:1210.1401}}
  (\bibinfo{year}{2012}).

\bibitem{brand2014macroscopic}
\bibinfo{author}{Brand, H.~R.} \& \bibinfo{author}{Pleiner, H.}
\newblock \bibinfo{title}{Macroscopic behavior of ferronematic gels and
  elastomers}.
\newblock \emph{\bibinfo{journal}{The European Physical Journal E}}
  \textbf{\bibinfo{volume}{37}}, \bibinfo{pages}{1--9} (\bibinfo{year}{2014}).

\bibitem{Nadzharyan2018316}
\bibinfo{author}{Nadzharyan, T.}, \bibinfo{author}{Kostrov, S.},
  \bibinfo{author}{Stepanov, G.} \& \bibinfo{author}{Kramarenko, E.}
\newblock \bibinfo{title}{Fractional rheological models of dynamic mechanical
  behavior of magnetoactive elastomers in magnetic fields}.
\newblock \emph{\bibinfo{journal}{Polymer}} \textbf{\bibinfo{volume}{142}},
  \bibinfo{pages}{316--329} (\bibinfo{year}{2018}).
\newblock \bibinfo{note}{Cited By 29}.

\bibitem{stolbov2011modelling}
\bibinfo{author}{Stolbov, O.~V.}, \bibinfo{author}{Raikher, Y.~L.} \&
  \bibinfo{author}{Balasoiu, M.}
\newblock \bibinfo{title}{Modelling of magnetodipolar striction in soft
  magnetic elastomers}.
\newblock \emph{\bibinfo{journal}{Soft Matter}} \textbf{\bibinfo{volume}{7}},
  \bibinfo{pages}{8484--8487} (\bibinfo{year}{2011}).

\bibitem{metsch2016numerical}
\bibinfo{author}{Metsch, P.}, \bibinfo{author}{Kalina, K.~A.},
  \bibinfo{author}{Spieler, C.} \& \bibinfo{author}{Kästner, M.}
\newblock \bibinfo{title}{A numerical study on magnetostrictive phenomena in
  magnetorheological elastomers}.
\newblock \emph{\bibinfo{journal}{Computational Materials Science}}
  \textbf{\bibinfo{volume}{124}}, \bibinfo{pages}{364--374}
  (\bibinfo{year}{2016}).
\newblock
  \urlprefix\url{https://www.sciencedirect.com/science/article/pii/S0927025616303822}.

\bibitem{guan-fem-vs-ezp-sufrace-roughness}
\bibinfo{author}{Guan, Y.} \emph{et~al.}
\newblock \bibinfo{title}{Surface morphology evolution of magnetorheological
  elastomers under magnetic fields: Simulation and experimental study}.
\newblock \emph{\bibinfo{journal}{Journal of Alloys and Compounds}}
  \textbf{\bibinfo{volume}{1051}}, \bibinfo{pages}{186044}
  (\bibinfo{year}{2026}).
\newblock
  \urlprefix\url{https://www.sciencedirect.com/science/article/pii/S092583882600112X}.

\bibitem{biller2019mesomechanical}
\bibinfo{author}{Biller, A.}, \bibinfo{author}{Stolbov, O.} \&
  \bibinfo{author}{Raikher, Y.~L.}
\newblock \bibinfo{title}{ in \textit{Mesomechanical response of a soft
  magnetic elastomer to {AC} magnetization}} (eds \bibinfo{editor}{Matveenko,
  V.~P.}, \bibinfo{editor}{Krommer, M.}, \bibinfo{editor}{Belyaev, A.~K.} \&
  \bibinfo{editor}{Irschik, H.}) \emph{\bibinfo{booktitle}{Dynamics and Control
  of Advanced Structures and Machines: Contributions from the 3rd International
  Workshop, Perm, Russia}} \bibinfo{pages}{39--48}
  (\bibinfo{publisher}{Springer}, \bibinfo{year}{2019}).

\bibitem{romeis2026beyond}
\bibinfo{author}{Romeis, D.}
\newblock \bibinfo{title}{Beyond the dipole approximation: A compact operator
  form to describe magnetizable many-body systems}.
\newblock \emph{\bibinfo{journal}{arXiv preprint arXiv:2604.13647}}
  (\bibinfo{year}{2026}).

\bibitem{pedro-elastomers-surface}
\bibinfo{author}{Sánchez, P.~A.}, \bibinfo{author}{Minina, E.~S.},
  \bibinfo{author}{Kantorovich, S.~S.} \& \bibinfo{author}{Kramarenko, E.~Y.}
\newblock \bibinfo{title}{Surface relief of magnetoactive elastomeric films in
  a homogeneous magnetic field: molecular dynamics simulations}.
\newblock \emph{\bibinfo{journal}{Soft Matter}} \textbf{\bibinfo{volume}{15}},
  \bibinfo{pages}{175--189} (\bibinfo{year}{2019}).
\newblock \urlprefix\url{http://dx.doi.org/10.1039/C8SM01850B}.

\bibitem{pedro-alla-inelastic-deformations}
\bibinfo{author}{Sánchez, P.~A.}, \bibinfo{author}{Gundermann, T.},
  \bibinfo{author}{Dobroserdova, A.}, \bibinfo{author}{Kantorovich, S.~S.} \&
  \bibinfo{author}{Odenbach, S.}
\newblock \bibinfo{title}{Importance of matrix inelastic deformations in the
  initial response of magnetic elastomers}.
\newblock \emph{\bibinfo{journal}{Soft Matter}} \textbf{\bibinfo{volume}{14}},
  \bibinfo{pages}{2170--2183} (\bibinfo{year}{2018}).
\newblock \urlprefix\url{http://dx.doi.org/10.1039/C7SM02366A}.

\bibitem{alla-flake-magneto-elastic-coupling}
\bibinfo{author}{Dobroserdova, A.} \emph{et~al.}
\newblock \bibinfo{title}{Magneto-elastic coupling as a key to microstructural
  response of magnetic elastomers with flake-like particles}.
\newblock \emph{\bibinfo{journal}{Soft Matter}} \textbf{\bibinfo{volume}{18}},
  \bibinfo{pages}{496--506} (\bibinfo{year}{2022}).
\newblock \urlprefix\url{http://dx.doi.org/10.1039/D1SM01349A}.

\bibitem{FORC_Dobroserdova_2020}
\bibinfo{author}{Dobroserdova, A.} \emph{et~al.}
\newblock \bibinfo{title}{Measuring forcs diagrams in computer simulations as a
  mean to gain microscopic insight}.
\newblock \emph{\bibinfo{journal}{Journal of Magnetism and Magnetic Materials}}
  \textbf{\bibinfo{volume}{501}} (\bibinfo{year}{2020}).

\bibitem{Dobroserdova_2023_PRE_FORC_SFD}
\bibinfo{author}{Dobroserdova, A.~B.}, \bibinfo{author}{Novak, E.~V.} \&
  \bibinfo{author}{Kantorovich, S.~S.}
\newblock \bibinfo{title}{Switching-field and first-order-reversal-curve
  distribution measurements in magnetic elastomers by molecular dynamics
  simulations: Accounting for polydispersity}.
\newblock \emph{\bibinfo{journal}{Phys. Rev. E}}
  \textbf{\bibinfo{volume}{107}}, \bibinfo{pages}{044606}
  (\bibinfo{year}{2023}).
\newblock \urlprefix\url{https://link.aps.org/doi/10.1103/PhysRevE.107.044606}.

\bibitem{dobroserdova2026influence}
\bibinfo{author}{Dobroserdova, A.~B.} \& \bibinfo{author}{Kantorovich, S.~S.}
\newblock \bibinfo{title}{Influence of particle-matrix coupling and magnetic
  anisotropy on the dynamic response of magnetic elastomers: A molecular
  dynamics study}.
\newblock \emph{\bibinfo{journal}{Journal of Molecular Liquids}}
  \textbf{\bibinfo{volume}{456}}, \bibinfo{pages}{129519}
  (\bibinfo{year}{2026}).
\newblock
  \urlprefix\url{https://www.sciencedirect.com/science/article/pii/S0167732226002904}.

\bibitem{rosensweig85a}
\bibinfo{author}{Rosensweig, R.~E.}
\newblock \emph{\bibinfo{title}{Ferrohydrodynamics}}
  (\bibinfo{publisher}{Cambridge Univ. Press}, \bibinfo{address}{Cambridge},
  \bibinfo{year}{1985}).

\bibitem{2007-bohlius}
\bibinfo{author}{Bohlius, S.}, \bibinfo{author}{Pleiner, H.} \&
  \bibinfo{author}{Brand, H.~R.}
\newblock \bibinfo{title}{Solution of the adjoint problem for instabilities
  with a deformable surface: Rosensweig and marangoni instability}.
\newblock \emph{\bibinfo{journal}{Phys. Fluids}} \textbf{\bibinfo{volume}{19}},
  \bibinfo{pages}{094103} (\bibinfo{year}{2007}).

\bibitem{Snarskii2019}
\bibinfo{author}{Snarskii, A.~A.}, \bibinfo{author}{Zorinets, D.},
  \bibinfo{author}{Shamonin, M.} \& \bibinfo{author}{Kalita, V.~M.}
\newblock \bibinfo{title}{Theoretical method for calculation of effective
  properties of composite materials with reconfigurable microstructure:
  Electric and magnetic phenomena}.
\newblock \emph{\bibinfo{journal}{Physica A: Statistical Mechanics and its
  Applications}} \textbf{\bibinfo{volume}{535}}, \bibinfo{pages}{122467}
  (\bibinfo{year}{2024}).

\bibitem{alexey-MMF2}
\bibinfo{author}{Ivanov, A.~O.} \& \bibinfo{author}{Kuznetsova, O.~B.}
\newblock \bibinfo{title}{Magnetic properties of dense ferrofluids: An
  influence of interparticle correlations}.
\newblock \emph{\bibinfo{journal}{Phys. Rev. E}} \textbf{\bibinfo{volume}{64}},
  \bibinfo{pages}{041405} (\bibinfo{year}{2001}).
\newblock \urlprefix\url{https://link.aps.org/doi/10.1103/PhysRevE.64.041405}.

\bibitem{gasper-magnetisation}
\bibinfo{author}{Glavan, G.}, \bibinfo{author}{Belyaeva, I.~A.},
  \bibinfo{author}{Ruwisch, K.}, \bibinfo{author}{Wollschläger, J.} \&
  \bibinfo{author}{Shamonin, M.}
\newblock \bibinfo{title}{Magnetoelectric response of laminated cantilevers
  comprising a magnetoactive elastomer and a piezoelectric polymer, in pulsed
  uniform magnetic fields}.
\newblock \emph{\bibinfo{journal}{Sensors}} \textbf{\bibinfo{volume}{21}}
  (\bibinfo{year}{2021}).
\newblock \urlprefix\url{https://www.mdpi.com/1424-8220/21/19/6390}.

\bibitem{gasper-magnetisation-zenodo}
\bibinfo{author}{Glavan, G.}, \bibinfo{author}{Belyaeva, I.},
  \bibinfo{author}{Ruwisch, K.}, \bibinfo{author}{Wollschläger, J.} \&
  \bibinfo{author}{Shamonin, M.}
\newblock \bibinfo{title}{Magnetoelectric response of laminated cantilevers
  comprising a magnetoactive elastomer and a piezoelectric polymer, in pulsed
  uniform magnetic fields} (\bibinfo{year}{2025}).
\newblock \urlprefix\url{https://doi.org/10.5281/zenodo.14652152}.

\bibitem{espresso-5}
\bibinfo{author}{Weeber, R.} \emph{et~al.}
\newblock \bibinfo{title}{ in \textit{Espresso, a versatile open-source
  software package for simulating soft matter systems}} \bibinfo{edition}{First
  edition} edn, (eds \bibinfo{editor}{Yáñez, M.} \& \bibinfo{editor}{Boyd,
  R.~J.}) \emph{\bibinfo{booktitle}{Comprehensive Computational Chemistry
  (First Edition)}} \bibinfo{pages}{578--601} (\bibinfo{publisher}{Elsevier},
  \bibinfo{address}{Oxford}, \bibinfo{year}{2024}).
\newblock
  \urlprefix\url{https://www.sciencedirect.com/science/article/pii/B9780128219782001033}.

\bibitem{espresso-5-zenodo}
\bibinfo{author}{Grad, J.-N.} \emph{et~al.}
\newblock \bibinfo{title}{Espresso} (\bibinfo{year}{2026}).
\newblock \urlprefix\url{https://doi.org/10.5281/zenodo.18791183}.

\bibitem{deniz-magnetizable-model}
\bibinfo{author}{Mostarac, D.}, \bibinfo{author}{Sánchez, P.~A.} \&
  \bibinfo{author}{Kantorovich, S.}
\newblock \bibinfo{title}{Characterisation of the magnetic response of
  nanoscale magnetic filaments in applied fields}.
\newblock \emph{\bibinfo{journal}{Nanoscale}} \textbf{\bibinfo{volume}{12}},
  \bibinfo{pages}{13933--13947} (\bibinfo{year}{2020}).
\newblock \urlprefix\url{http://dx.doi.org/10.1039/D0NR01646B}.

\bibitem{Griffiths2023}
\bibinfo{author}{Griffiths, D.~J.}
\newblock \emph{\bibinfo{title}{Introduction to Electrodynamics}}
  (\bibinfo{publisher}{Cambridge University Press}, \bibinfo{year}{2023}).

\bibitem{ITIM-original}
\bibinfo{author}{P{\'a}rtay, L.~B.}, \bibinfo{author}{Hantal, G.},
  \bibinfo{author}{Jedlovszky, P.}, \bibinfo{author}{Vincze, {\'A}.} \&
  \bibinfo{author}{Horvai, G.}
\newblock \bibinfo{title}{A new method for determining the interfacial
  molecules and characterizing the surface roughness in computer simulations.
  application to the liquid–vapor interface of water}.
\newblock \emph{\bibinfo{journal}{Journal of Computational Chemistry}}
  \textbf{\bibinfo{volume}{29}}, \bibinfo{pages}{945--956}
  (\bibinfo{year}{2008}).
\newblock
  \urlprefix\url{https://onlinelibrary.wiley.com/doi/abs/10.1002/jcc.20852}.

\bibitem{steinhardt83bond}
\bibinfo{author}{Steinhardt, P.~J.}, \bibinfo{author}{Nelson, D.~R.} \&
  \bibinfo{author}{Ronchetti, M.}
\newblock \bibinfo{title}{Bond-orientational order in liquids and glasses}.
\newblock \emph{\bibinfo{journal}{Phys. Rev. B}} \textbf{\bibinfo{volume}{28}},
  \bibinfo{pages}{784--805} (\bibinfo{year}{1983}).
\newblock \urlprefix\url{https://link.aps.org/doi/10.1103/PhysRevB.28.784}.

\bibitem{lechner08accurate}
\bibinfo{author}{Lechner, W.} \& \bibinfo{author}{Dellago, C.}
\newblock \bibinfo{title}{Accurate determination of crystal structures based on
  averaged local bond order parameters}.
\newblock \emph{\bibinfo{journal}{The Journal of Chemical Physics}}
  \textbf{\bibinfo{volume}{129}}, \bibinfo{pages}{114707}
  (\bibinfo{year}{2008}).
\newblock \urlprefix\url{https://doi.org/10.1063/1.2977970}.

\bibitem{belyaeva2017magnetodielectric}
\bibinfo{author}{Belyaeva, I.~A.}, \bibinfo{author}{Kramarenko, E.~Y.} \&
  \bibinfo{author}{Shamonin, M.}
\newblock \bibinfo{title}{Magnetodielectric effect in magnetoactive elastomers:
  Transient response and hysteresis}.
\newblock \emph{\bibinfo{journal}{Polymer}} \textbf{\bibinfo{volume}{127}},
  \bibinfo{pages}{119--128} (\bibinfo{year}{2017}).

\bibitem{paper1-data-zenodo}
\bibinfo{author}{P.~A.~Santos, J.}
\newblock \bibinfo{title}{Simulation data} (\bibinfo{year}{2026}).
\newblock \urlprefix\url{https://doi.org/10.5281/zenodo.22872568}.

\bibitem{mahdi-data-E-and-R}
\bibinfo{author}{Hasanzade, M.} \emph{et~al.}
\newblock \bibinfo{title}{Experimental characterization of magnetically tunable
  mechanical and surface properties of isotropic magnetoactive elastomers}
  (\bibinfo{year}{2026}).
\newblock \urlprefix\url{https://doi.org/10.17632/7z7966yfyx.4}.

\bibitem{pressomancy-github}
\bibinfo{author}{Mostarac, D.} \& \bibinfo{author}{P.~A.~Santos, J.}
\newblock \bibinfo{title}{Pressomancy} (\bibinfo{year}{2026}).
\newblock \urlprefix\url{https://github.com/stekajack/pressomancy.git}.

\bibitem{paper1-code-repo}
\bibinfo{author}{P.~A.~Santos, J.}
\newblock \bibinfo{title}{Simulation and analysis code} (\bibinfo{year}{2026}).
\newblock
  \urlprefix\url{https://github.com/juliopas/codebase_and_analysis_MHvsMS.git}.

\end{thebibliography}

\section*{Acknowledgements}\label{sec:ack}


This research was funded by the European Union's Horizon Europe research and innovation programme under Marie Sk\l{}odowska-Curie Actions Doctoral Network MAESTRI (grant agreement no. 101119614). Funded by the European Union. Views and opinions expressed are those of the authors only and do not necessarily reflect those of the European Union or the European Research Executive Agency; neither the European Union nor the granting authority can be held responsible for them.
R.K. and M.S. acknowledge financial support from the Deutsche Forschungsgemeinschaft (DFG; project number 524921900).
The computational results were achieved using the Austrian Scientific Computing (ASC) infrastructure.
C.D. and M.S. thank Evonik Operations GmbH, Specialty Additives (Geesthacht, Germany) and BASF SE (Ludwigshafen am Rhein, Germany) for supplying materials.

\section*{Author contributions}\label{sec:contrib}


J.P.A.S. developed the magnetizable-particle model, performed and analyzed the simulations, analyzed the experimental data and wrote the manuscript.
S.K. conceived and supervised the computational study, contributed to the interpretation of the results, revised the manuscript and acquired funding.
M.S. supervised the experimental work, contributed to the interpretation of the results, revised the manuscript and acquired funding.
R.K. performed the surface-topography measurements with M.H., proposed the comparison of samples at equal zero-field stiffness and revised the manuscript.
C.D. prepared the samples.
M.H. performed the surface-topography and stiffness measurements and processed the surface-topography data.
C.D. and M.H. wrote the experimental Methods.
A.K. assisted with the stiffness measurements setup and the processing of the experimental data.
S.N.G. supervised the experimental work and acquired funding
All authors reviewed and approved the submitted version.

\section*{Competing interests}\label{sec:competing}

The authors declare no competing interests.


\begin{table*}[t]
    \centering
    \caption{Conversion from the non-dimensional field $\xi$ to the magnetic flux density in SI units and in simulation units. The temperatures used for the experimental and simulation values are defined in the \textit{Methods} (``Mapping to physical units'').}
    \label{tab: xi-to-B}
    \begin{tabular}{@{}lccccccc@{}}
        \toprule
        $\xi\ (\times10^{9})$ & $0.06$ & $0.09$ & $0.12$ & $0.19$ & $0.62$ & $1.50$ & $1.86$ \\
        \midrule
        $B$ (\si{\milli\tesla}) & $10$ & $15$ & $20$ & $32$ & $106$ & $256$ & $317$ \\
        $B$ (sim. units) & $1$ & $1.5$ & $2$ & $3$ & $10$ & $25$ & $30$ \\
        \botrule
    \end{tabular}
\end{table*}

\end{document}